\documentclass[12pt,peerreview,draftclsnofoot,letterpaper,oneside,onecolumn]{IEEEtran}
\IEEEoverridecommandlockouts
\usepackage{}

\usepackage{amsfonts,amsmath,amssymb}
\usepackage{cite}
\usepackage{graphicx}
\usepackage{epstopdf}
\usepackage{algorithm}
\usepackage{algorithmic}
\usepackage{mathbbold}
\usepackage{mathrsfs}
\usepackage{xcolor}
\usepackage{stfloats}
\usepackage{cases}

\usepackage{amsfonts,amsmath,amssymb}
\usepackage{cite}
\usepackage{graphicx}
\usepackage{url}

 \newtheorem{lemma}{Lemma}
  \newtheorem{proposition}{Proposition}
 \newtheorem{corollary}{Corollary}

\usepackage{bm}
\usepackage{multirow}

\begin{document}

\title{ Adaptively Directional Wireless Power Transfer for Large-scale Sensor Networks}

%Adaptively Directional

\author{Zhe~Wang,~\IEEEmembership{Member,~IEEE}, Lingjie Duan,~\IEEEmembership{Member,~IEEE}, and Rui Zhang,~\IEEEmembership{Senior Member,~IEEE}
 \thanks{
Part of this work is presented at the IEEE Global Communications Conference (GLOBECOM 2015), San Diego, CA, USA, Dec. 6-10, 2015.
}
 \thanks{
 Z. Wang and L. Duan are with Pillar of Engineering Systems and Design, Singapore University of Technology and Design,
  Singapore (e-mail: zhe$\_$wang@sutd.edu.sg; lingjie$\_$duan@sutd.edu.sg).
}
\thanks{R. Zhang is with the Department of Electrical and Computer Engineering,
National University of Singapore, Singapore; and also with the
Institute for Infocomm Research, A*STAR, Singapore (e-mail:
elezhang@nus.edu.sg)}
}

\maketitle

\maketitle
\begin{abstract}
Wireless power transfer (WPT) prolongs the lifetime of wireless sensor network by providing sustainable power supply to the distributed sensor nodes (SNs) via electromagnetic waves.
To improve the energy transfer efficiency in a large WPT system, this paper proposes an adaptively directional WPT (AD-WPT) scheme, where the power beacons (PBs) adapt the energy beamforming strategy to SNs' locations by concentrating the transmit power on the nearby SNs within the efficient charging radius.
With the aid of stochastic geometry, we derive the closed-form expressions of the distribution metrics of the aggregate received power at a typical SN and further approximate the complementary cumulative distribution function using Gamma distribution with second-order moment matching.
To design the charging radius for the optimal AD-WPT operation, we exploit the tradeoff between the power intensity of the energy beams and the number of SNs to be charged.
Depending on different SN task requirements, the optimal AD-WPT can maximize the average received power or the active probability of the SNs, respectively. It is shown that both the maximized average received power and the maximized sensor active probability increase with the increased deployment density and transmit power of the PBs, and decrease with the increased density of the SNs and the energy beamwidth.
Finally, we show that the optimal AD-WPT can significantly improve the energy transfer efficiency compared with the traditional omnidirectional WPT.
\end{abstract}
%In this paper, we propose an adaptively directional wireless power transfer (AD-WPT) scheme for a sensor network, where the power beacons (PBs) charge the nearby SNs that are within the charging regions by adapting the energy beamforming to the sensor locations.

\newpage

\section{Introduction}
%, which . self-sustainable...

%why large solar panels? --- increase the energy harvesting efficiency
%why large battery? --- store the robust energy from ambient environment.
%However, SNs are small devices. ---need WPT.
%Harvesting energy from natural environment is may not be an ideal choice for WSN since the small-size SNs may not be able to deploy energy harvesting devices, e.g., solar panels, or large battery storage modules that used to store the robust energy from the ambient environment.

Wireless sensor networks (WSNs) consist of small-size, low-power and distributed sensor nodes (SNs) to monitor physical or environmental conditions \cite{IFsensor}. WSNs are often required to operate for long periods of time, but the network lifetime is constrained by the limited battery capacity and costly battery replacement at SNs. To extend the network lifetime, it is desirable to recharge the SNs in an undisruptive and energy efficient way.

%The recent advances of RF-based wireless power transfer (WPT) \cite{RF} enables power beacons (PBs)
%with constant power supplies (e.g., batteries)
%to charge the SNs via electromagnetic (EM)  waves, which provides a controllable and sustainable power supply to the WSN %\cite{sensor1,sensor2,sensor3}.

RF-enabled wireless power transfer (WPT) \cite{Bi} provides a controllable and sustainable power supply to sensor network by charging SNs via electromagnetic (EM) waves \cite{sensor1,sensor2,sensor3}.
There are mainly two types of WPT: omnidirectional WPT and directional WPT. For omnidirectional WPT, the energy transmitter or so-called  power beacon (PB) broadcasts the EM waves equally in all directions regardless of the locations of the energy receivers. According to the law of conservation of energy, the energy radiated in the direction of energy receivers accounts for only a small fraction of the total radiated power.
%Since the EM waves fade rapidly over distance, charging via omnidirectional WPT for a target received power may require excessive high transmit power, which may not be energy efficient.
Since the EM waves fade rapidly over distance, it may require excessively high transmit power to charge an energy receiver via omnidirectional WPT, which may not be energy efficient.
%Since the EM waves fade rapidly over distance, excessive high transmit power may be needed to achieve the target received power. omnidirectional WPT energy inefficient.
%In contrast, directional WPT is more energy efficient for a green wireless charging system.
%reduce the carbon dioxide emissions
%The directional WPT is more to reduce the CO$_2$ emissions
% Add a sentence about green communication.....
In contrast, for  directional WPT with antenna arrays, the PB concentrates the radiated energy in the directions of the energy receivers, i.e., via energy beamforming, which enhances the power intensity in the intended directions. The energy transfer efficiency is thus improved with the consequent reduction of transmit power to reach the target received power.
%, which makes directional WPT a better choice for a  green power transfer system.
%With directional WPT, the WPT system is green due to the reduced carbon dioxide emission for the required received power at receivers.
%with the consequent reduction of carbon dioxide emission for the required received energy at receivers.

%For each direction, the power intensity is low since the energy radiated in this direction accounts for only a small fraction of the total radiated power.
Most of the literature on directional WPT (see \cite{Bi} and references therein) has focused on point-to-point and point-to-multipoint transmissions. For a large-scale  WSN, the SNs are often in large quantities and are usually  distributed with random locations. There are two main challenges in the design of directional WPT for a large-scale network. On the \emph{PB-side}, it is challenging to adapt the energy beamforming strategy to the random locations of the SNs, e.g., to decide which SNs to serve, how many beams to generate and the beamwidth of each beam, etc. On the \emph{SN-side}, it is difficult  to analyze the aggregate received power from a large number of PBs in the network, where  the radiation directions and energy intensity may vary for each PB.

In this paper, we aim at tackling the above two challenges.
The paper structure  and main contributions are given as follows.
\begin{itemize}
\item \textit{Energy-efficient AD-WPT scheme to power a large-scale sensor network}:
To address the PB-side challenge, we propose an adaptively directional WPT (AD-WPT) scheme in a large-scale sensor network in Section II, where the energy beamforming strategy of the PBs is adaptive to the nearby SN locations that are within the energy-efficient charging radius. To deal with
the tradeoff between the power intensity of the energy beams and the number of SNs served by each PB, we design the charging radius to achieve optimal AD-WPT for different performance targets, i.e., average power maximization or active probability maximization.
\item \textit{Analysis of harvested  power using stochastic geometry}:
To address the SN-side challenge, in Section III, we successfully derive the closed-form expressions of the distribution metrics, e.g., Laplace transform, mean and variance, of the aggregate received power at a typical SN from the large-scale PB network using the tools of stochastic geometry \cite{weber,martinbook,stochastic}. The complementary cumulative distribution function (CCDF) of the received power is also analyzed. As it is difficult to obtain the analytical CCDF expression, we further  approximate it using Gamma distribution with second-order moment matching.
%The average received power is a unimodal function of charging radius since there exists a tradeoff between the coverage of the charging region and energy concentration.
\item \textit{Optimal AD-WPT for average power maximization}:
In flexible-task WSN, the SNs operate in a cooperative manner on power adaptive sensing tasks.
%some SNs are allowed to receive higher power than other SNs to perform more sensing or coordination tasks.
To achieve the optimal AD-WPT, we design the optimal charging radius to maximize the average received power of the SNs in Section IV.
We show that the maximized average received power increases with the increased PB power and density, while it decreases with the increased energy beamwidth and SN density.
%The optimal charging radius is decreasing with the increased energy beamwidth and SN density and it is regardless of the PB power and density.
In addition, the optimal AD-WPT greatly improves the average received power compared with the traditional omnidirectional WPT, especially when PB power/density is high.
\item \textit{Optimal AD-WPT for active probability maximization}:
In equal-task WSN, the SNs operate in an independent manner on equal quantity of sensing tasks, where an SN is active if its received power is larger than the operational power threshold.
%each SN has the same quantity of sensing tasks with a minimum power requirement. Certain achieve the sensing diversity.
%An SN is inactive if the aggregate received power at a SN from all PBs falls below the threshold.
To achieve the  optimal AD-WPT, we design the optimal charging radius to maximize the active probability of the SNs in Section V.
%It shows that the maximized sensor active probability also increases with the increased PB power and density, while it decreases with the increased energy beamwidth and SN density. Moreover, t
It shows that the optimal AD-WPT can enhance the sensor active probability compared with omnidirectional WPT, especially when the PB power/density is not high.
\end{itemize}
In Section VI, the numerical results are shown and discussed. Finally, conclusions are drawn in Section VII.

\subsection{Related Literature}
%Some related works are listed as follows.
Omnidirectional WPT has been studied recently in \cite{p2p1,p2p2,CR,YL}. In \cite{p2p1}, a point-to-point omnidirectional WPT is investigated, where the receiver utilizes part of the harvested energy for decoding the information in the received signal. In \cite{p2p2}, the downlink energy transfer in a broadcast network is studied for throughput maximization. In \cite{CR}, a stochastic geometry based model is considered for a cognitive radio network, where the secondary transmitters harvest RF energy from the nearby primary transmitters. \cite{YL} investigates the downlink energy transfer in a large-scale wireless network  by considering finite and infinite battery capacity.

The directional WPT has been addressed in \cite{beamp2p,beambroadcast,beambroadcast2,KB2}. In \cite{beamp2p}, energy beamforming is studied in a broadcast network where the transmitter steers the energy beams towards the receivers to maximize their received power. In \cite{beambroadcast} and \cite{beambroadcast2}, energy beamforming is designed  in a MIMO broadcast network jointly with information beamforming, where the transmitter adjusts the beam weights to maximize the received power and information rate at different receivers. In \cite{KB2}, each mobile node in a cellular network is charged by its nearest PB via energy beamforming. For the simplicity of analysis, only the received power from the nearest PB is considered and the received energy from all other PBs is omitted in \cite{KB2}.

To the best of our  knowledge, this paper is the first study of directional WPT by using  adaptive energy beamforming for a large-scale network and the resulting aggregate received power from all PBs with AD-WPT is rigorously characterized. With the proposed AD-WPT scheme, the energy transfer efficiency in the large-scale network can be greatly enhanced compared with the traditional omnidirectional WPT.

\section{System Model}

We consider a wireless charging network  as shown in Fig. 1, where a PB network wirelessly charges an SN network via energy beamforming. Each PB radiates EM waves with wavelength $\nu$ using transmit power $P_p$.
The PBs and SNs follow two independent homogeneous Poisson Point Processes (PPPs) $\Phi_p=\{X_i\}$ and $\Phi_{s}=\{Y_j\}$ with density $\lambda_p$ and $\lambda_s$, respectively, where $X_i$ and $Y_j$ represent the coordinates of the PBs and SNs in $\mathbb{R}^2$ plane.
%It is assumed that the locations of PBs and SNs remain the same for each time slot while they may vary over different slots.
%No battery storage module is assumed at SNs.

% \footnotemark. Each PB radiates EM waves with a fixed transmit power $P_p$.

%\footnotetext{The battery storage of the SNs is not discussed in this paper since the complicated analysis of  is not our focus. }

In the following, we first propose a power transfer scheme with adaptive energy beamforming and then discuss the power intensity in the directions of the energy beams.
\begin{figure}[t!]
    \begin{center}
        \includegraphics[width=0.6\columnwidth]{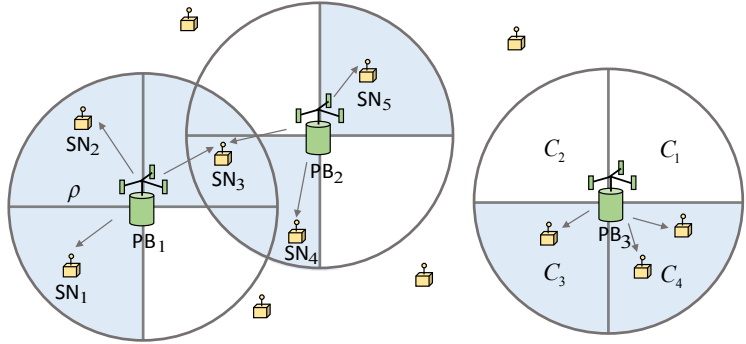}
        \caption{System model of AD-WPT (illustrative example of $N=4$). The circular areas  with radius $\rho$ are the charging regions of the PBs. The shaded sectors in the charging regions are the active sectors of the PBs.}
        %The circular areas are the charging regions with radius $\rho$. The shaded area are the active sectors of the PBs.}
    \end{center}
    \label{system}
\end{figure}

\subsection{AD-WPT Scheme}
Due to the fast attenuation of the radio power over the distance, it is more energy efficient for the PBs to focus the energy to charge the nearby SNs. With antenna arrays, a PB is able to form an energy beam in a certain direction or generate multiple beams simultaneously towards different directions \cite{antennabook}.
In this subsection, we propose an AD-WPT scheme where the PBs adapt the beamforming strategy to the random locations of the SNs.

To decide which SNs to charge, we define \emph{charging region} as a circular region centered at each PB with \emph{charging radius} $\rho$, as shown in Fig. 1.
%Note the charging radius is designed to decide ..., the power radiation of PB all distance, not limited to ...
%We assume the PBs is aware of the existence of the SN (at least one SN) within its charging region.
Each charging region is divided into $N$ equal sectors $C_1,\cdots,C_N$, where $N$ is usually a small positive integer due to physical constraint of antenna design.
%We assume the PBs have the location information of the nearby SNs within its charging region.
We consider that a PB is aware of the existence of the SNs inside each of its sectors, e.g., via the SN feedback over control channels.
A sector is considered to be active if at least one SN falls into this sector. Denote $M$ as the random number of active sectors of a PB, e.g., PB$_i$, where $0\leq{M}\leq{N}$. The adaptive beamforming strategy of PB$_i$ is given as follows.
\begin{itemize}
\item SN's absence in charging region: If no sector of PB$_i$ is active ($M=0$), PB$_i$ works as an omnidirectional antenna that radiates energy equally in all directions (to help power SNs outside the charging region).
\item SN's presence in charging region: If at least one sector of PB$_i$ is active ($M\geq1$), PB$_i$ generates $M$ equal-power energy beams in the directions of the $M$ active sectors.
%Assume the side lobes are negligible.
\end{itemize}
We use equal power allocation among the energy beams of a PB for the ease of analysis. In Section VI-C, we will show that equal power allocation
is descent as compared with some other unequal allocation choices.
%is almost as good as adaptive power allocation which allocates power according to the number of SNs in each sector.

From an SN's point of view, the received power from the PBs is discussed as follows.
\begin{itemize}
\item Inside charging region (or within radius $\rho$): An SN can be intentionally and efficiently  charged by one or more PBs whose charging regions cover its location.
\item Outside charging region (or beyond radius $\rho$): When an SN is located outside the charging regions of the PBs, the SN still  receives RF energy from the PBs if it is aligned with the energy radiation directions of the PBs.
\end{itemize}

We further explain the proposed AD-WPT scheme with the example of $N=4$ in Fig. 1. It is observed that PB$_1$ detects three nearby sensors, i.e., SN$_1$, SN$_2$ and SN$_3$, which fall into three out of four  sectors of its charging region. As a result, PB$_1$ adaptively generates three energy beams in the directions of north-east, north-west and south-west to directionally charge the three sensors. At the same time, PB$_2$ detects three sensors, i.e., SN$_3$, SN$_4$ and SN$_5$, which fall into two sectors of its charging region, and thus two adaptive energy beams are generated towards these SNs. In particular, notice that SN$_3$, which is within the overlapping area of the charging regions of PB$_1$ and PB$_2$, is thus intentionally charged by the two PBs at the same time. SN$_1$ is intentionally charged by PB$_1$ while it also receives energy from PB$_2$ and PB$_3$ since its location  is aligned with the south-west energy radiation directions the two PBs.

\subsection{Antenna Gain under AD-WPT}
%\begin{figure}[t!]
%    \begin{center}
%        \includegraphics[width=1.00\columnwidth]{solid.eps}
%        \caption{Horizontal cross section of $N$ solid angles ($N=4$ and $N=8$).}
%    \end{center}
%    \label{solid}
%\end{figure}
%\cite{antennabook, pathloss}
When a PB is directional, the power intensity in the directions of energy beams improves compared with the case when the PB is omnidirectional. The ratio of power intensity between  directional and omnidirectional antenna is defined as the gain of directional antenna $G$ ($G\geq1$) \cite{antennabook}. In the unintended directions of the directional PB, the power intensity is zero. In the following, we evaluate $G$ given that $M$ out of $N$ sectors of the PB are active.

If none of the sectors of a PB is active ($M=0$), as discussed, the PB behaves as an omnidirectional antenna with the uniform gain in all directions, i.e.,
\begin{align}
G_M=1,~\text{for}~M=0 \label{Gm0}.
\end{align}
If $M$ out of $N$ sectors of the PB are active ($M\geq1$), the PB forms $M$ ($M\leq{N}$) energy beams with equal power in the direction of each beam.\footnotemark~By  the law of conservation of energy, the total radiated power for directional and omnidirectional antenna is the same. Since the directional antenna concentrates the energy from the directions of $N$ sectors into $M$ sectors, the power intensity in the intended directions becomes $N/M$ times of that of the omnidirectional antenna. Therefore, given $M$ energy beams at the PB, the antenna gain in the direction of each energy beam is approximated as
\begin{align}
G_M={N}/{M},~\text{for}~M=1,\cdots,N. \label{Gmn}
\end{align}
From \eqref{Gm0} and \eqref{Gmn}, we see that the proposed AD-WPT is equivalent to the omnidirectional WPT with uniform gain when $M=0$ or $M=N$.

\footnotetext{For simplicity, we assume the side lobes are negligible and the radiated energy is uniformly distributed across each energy beam.}

%As $\rho$ increases from $0$ to $\infty$, it becomes more likely to include more SNs inside the charging region, which activates more sectors of the PB. The directional antenna gain $G_M$ increases from $G_0=1$ to $G_1=N$ and then decreases to $G_2=\frac{N}{2}$, $G_3=\frac{N}{3}$ until $G_N=1$. Since the power gain increases and then decreases with the increased $\rho$, the optimal charging radius is crucial in the AD-WPT design and will be analyzed in Section IV and Section V.

%As $\rho\rightarrow{0}$,  the AD-WPT scheme is equivalent to the omnidirectional WPT with gain $G_0=1$ since no SN is inside the charging region. As the increase of $\rho$, it is more likely to include more SNs inside the charging region, which activates more sectors of the PB. The antenna gain increases from $G_0=1$ to $G_1=N$ and then decreases to $G_2=\frac{N}{2}$, $G_3=\frac{N}{3}$ until $G_N=1$. As $\rho\rightarrow{\infty}$, it is likely that all $N$ sectors have SNs and the AD-WPT scheme is again equivalent to the isotropic WPT with $G_N=1$.

%In Fig. 2, given the PB density $\lambda_p$, SN density $\lambda_s$ and number of sectors per PB $N$, we plot the simulation results of the percentage of PBs that radiate with gain $G_M$ (for $M=0,1,\cdots,N$) and the percentage of PBs that radiate energy towards a typical SN, i.e., SN$_0$.
%For each charging region radius $\rho$, there is a gain $G_M$ that most PBs are likely to radiate with.
The antenna gains and number of energy beams of the PBs are related to the charging radius $\rho$.
As $\rho\rightarrow{0}$,  no SN is inside the charging regions ($M=0$) and all PBs radiate energy in $N$ directions with gain $G_0=1$ as omnidirectional WPT.  As the increase of $\rho$, more sectors of the PB are likely to be activated due to the increased number of SNs inside the charging region. The number of beams that most PB radiate with decreases from $N$ to $1$ sharply  and then increases from $1$, $2$, $\cdots$, to $N$. The corresponding antenna gain increases from $G_0=1$ to $G_1=N$ and then decreases from $G_1=N$, $G_2=\frac{N}{2}$, $\cdots$, to $G_N=1$.
%The power intensity of each beam increases when the PB focus energy into less beams.
As $\rho\rightarrow{\infty}$, AD-WPT is again equivalent to omnidirectional WPT with $G_N=1$ in all $N$ directions.
As we can see, there is a tradeoff between the antenna gain $G_M$ and the number of beams of the PBs.
%In other words, there is a tradeoff between the power intensity of the PBs and the number of SNs to be served by the PBs.
When the PB concentrates  energy on fewer beams, the power intensity of each beam increases  but at the cost of charging fewer SNs.
%The number of beams that most PBs radiate with increases and then decreases with the increased $\rho$, but the percentage of the PBs that radiate energy towards SN$_0$ decreases and then increases with $\rho$.
To address the above tradeoff, the optimal charging radius is crucial in the AD-WPT design and will be analyzed in Section IV and Section V for different SN  network requirements.
%\begin{figure}[t!]
%    \begin{center}
%        \includegraphics[width=1\columnwidth]{GM3.eps}
%        \caption{Percentage of PBs ($\lambda_p=0.1$, $\lambda_s=0.2$ and $N=4$).}
%    \end{center}
%\end{figure}

%From the PB's perspective, there is a tradeoff between the radiated power intensity and the number of SNs to be intentionally recharged. When the PB intentionally recharges more SNs, the received power intensity at each SN inside its charging region decreases due to the increased number of beams. By choosing a proper charging radius, the energy efficiency of the AD-WPT can be maximized.
 %This tradeoff also explains the existence of an optimal charging radius.
% note: the above discussion is partially correct. When rho is very small, the intensity increase with the increased rho (more SNs), but later it decreases with the increased rho. the tradeoff between "no. of SNs" and "intensity" only exists in the regime of large rho.

\section{Characterization of SNs' Received Power Using Stochastic Geometry }
In this section, we first study the aggregate received power at a typical SN from all PBs and then use stochastic geometry to analyze the distribution of the received power.

Consider a typical sensor node SN$_0$ at the origin and an arbitrary PB$_i$ at location $X_i$.
If PB$_i$ radiates energy with gain $G_M$ (for $M=0,1,\cdots,N$) towards SN$_0$, the received power at SN$_0$ from PB$_i$ is \cite{Goldsmithbook}
\begin{align}
P_{s}^{i}=P_{p}G_M\sigma\left[\max\left({\|X_i\|}/{d_0},1\right)\right]^{-\alpha},\label{Psi}
\end{align}
where $P_p$ is the transmit power of PB$_i$, $\alpha$ is the path loss exponent, $\sigma$ is a unitless constant depending on the receiver energy convention efficiency, antenna characteristics and average channel attenuation.\footnotemark~The Euclidian distance between PB$_i$ and SN$_0$ is represented by $\|{X_i}\|$, and $d_0$ is a reference distance for the antenna far field.
%To better characterize the effect of user locations, we assume that SN's mobility scale is much larger than the fading coherence time.
The received power from each PB is taken by averaging  over the short-term fading.
We adopt the non-singular path loss model \cite{martinbook} to avoid $\left[{\|X_i\|}/{d_0}\right]^{-\alpha}>1$ for ${\|X_i\|}<{d_0}$. Without of the loss of generality, we use $d_0=1$ throughout the paper.

\footnotetext{For empirical approximation, $\sigma$ is sometimes set to free-space path loss at distance $d_0$ assuming omnidirectional antennas, i.e., $\sigma=20\log_{10}\frac{\nu}{4\pi{d_0}}$ dB \cite{Goldsmithbook}, where $\nu$ is the wavelength of the radio waves.}

%Denote $\mathbbold{1}\left(\text{SN$_0$ receives $G_M$ from PB$_i$}\right)$ the event that SN$_0$ receives $G_M$ from PB$_i$.

Equation \eqref{Psi} holds if PB$_i$ radiates energy with gain $G_M$ towards SN$_0$, where $G_M$ is given in \eqref{Gm0} or \eqref{Gmn} depending on the number $M$ of active sectors of PB$_i$. By considering all PBs in the network, the aggregate received power at SN$_0$ is
\begin{align}
P_s&=\sum\limits_{X_i\in\Phi_p}P_s^{i}\mathbbold{1}\left(\text{SN$_0$ receives energy from PB$_i$ with $G_M$}\right).
\end{align}
%For $N>1$, the PBs may become directional radiators.
The indicator function equals one if both the following conditions are satisfied:
\begin{itemize}
\item Condition 1: PB$_i$ has $M$ active sectors;
\item Condition 2: SN$_0$ is in one of the $M$ radiation directions of PB$_i$ given PB$_i$ has $M$ active sectors.
\end{itemize}
We see that both conditions are related to the distance between SN$_0$ and PB$_i$.
If SN$_0$ is inside the charging region of PB$_i$, PB$_i$ generates at least one beam towards SN$_0$ ($M\geq1$).
%The value of $M$ also depends on other SNs' locations inside the charging radius of PB$_i$.
If SN$_0$ is outside the charging region of PB$_i$, SN$_0$ may not be in the radiation direction of PB$_i$ and $M$ may vary from $0$ to $N$.

According to the distance between PB$_i$ and SN$_0$, we classify the PBs into two groups: \textit{near} PBs with $\|X_i\|\leq{\rho}$, and \textit{far} PBs with $\|X_i\|>{\rho}$. We draw an equivalent charging region centered at SN$_0$ with radius $\rho$ and denote ${b}(o,\rho)$ and $\overline{b(o,\rho)}$ as the regions inside and outside this charging region, respectively. We define two indicator functions $\theta_n^M$ and $\theta_f^M$ to describe the events that SN$_0$ receives power from the PB with $G_M$ conditioned on this PB is a near PB or far PB, respectively, i.e.,
\begin{align}
\theta_n^M=\mathbbold{1}\left[\text{SN$_0$ receives energy from PB$_i$ with $G_M$}\mid\|X_i\|\leq{\rho}\right]
\end{align}
and
\begin{align}
\theta_f^M=\mathbbold{1}\left[\text{SN$_0$ receives energy from PB$_i$ with $G_M$}\mid\|X_i\|>{\rho}\right],
\end{align}
where subscripts $n$ and $f$ denote the near and far PBs and superscript $M$ denotes the number of active sectors of the PB.

We denote $P_{s,n}$ as the aggregate received power from the near PBs and $P_{s,f}$ as the aggregate received power from the far PBs that radiate energy towards SN$_0$.
By summing them up, we rewrite $P_s$ as
\begin{align}
P_s=P_{s,n}+P_{s,f},\label{dirins}
\end{align}
where
\begin{align}
P_{s,n}=P_p\sigma\sum\limits_{X_i\in\Phi_p\bigcap{b}(o,\rho)}{G}_M\theta_n^M\left[\max\left(\|X_i\|,1\right)\right]^{-\alpha}
\end{align}
and
\begin{align}
P_{s,f}&=P_p\sigma\sum\limits_{X_i\in\Phi_p\bigcap{\overline{b(o,\rho)}}}{G}_M\theta_f^M\left[\max\left(\|X_i\|,1\right)\right]^{-\alpha}.
\end{align}
As a special case of $N=1$, all PBs are omnidirectional radiators with gain of $1$. The aggregate received power at SN$_0$ from all omnidirectional PBs is
\begin{align}
P_s^{omni}&=P_p\sigma\sum\limits_{X_i\in\Phi_p}\left[\max\left(\|X_i\|,1\right)\right]^{-\alpha}.\label{Psiso}
\end{align}

%first $\theta_n^M$ and $\theta_f^M$, and then $P_s$...

%In the following discussions, we firstly derive the probability that SN$_0$ receives energy from PB$_i$ with antenna gain $G_M$ given PB$_i$ is a near PB and far PB, respectively. And then we use the conditional probabilities to derivation of the Laplace transform of the received power distribution.

To fully characterize the received power distribution, we usually use Laplace transform, which is however, difficult to  be derived  directly from \eqref{dirins}. In the conditional events of $\theta_n^M$ and $\theta_f^M$, the gain $G_M$ of PB$_i$ is  also related to the locations of other nearby SNs of PB$_i$ which are unknown. Moreover, since $G_M$ vary for each PB, the PBs that radiate power with $G_M$ towards SN$_0$ can be regarded as a heterogeneous network for which the Laplace transform is hard to characterize. In the following discussions, we use an alternative method by taking the privilege of the independent thinning \cite{stochastic} of the network. For the near PBs and the far PBs, respectively, we thin the heterogeneous network into multiple homogeneous networks with certain probabilities, where in each homogeneous network the PBs radiate energy towards SN$_0$ with the same gain $G_M$.
We have $M=1,\cdots,N$ for the near PBs and $M=0,1,\cdots,N$ for the far PBs.
%The thinning processes are given for the near PBs and far PBs, respectively.
After analyzing the Laplace transform of the received power distribution in each homogeneous network, we finally derive the distribution metrics of the aggregate received power  from all PBs at SN$_0$.

\subsection{Power Reception Probability given PB Location}
First, we derive the thinning probabilities of the near PBs and the far PBs.
As discussed previously, SN$_0$ receives power from PB$_i$ with gain $G_M$ if both Conditions 1 and 2  are satisfied. As for Condition 1, PB$_i$ transmits with gain $G_M$ if it has $M$ active sectors. We derive the active probability of each sector as follows. As SNs follow PPP with density $\lambda_s$, the number of SNs inside a charging region is a Poisson random variable with mean ${\lambda_s\pi{\rho^2}}$. When the charging region is equally partitioned into $N$ sectors, the number of SNs inside one of these $N$ sectors is also a Poisson random variable, denoted by $l$, with mean ${\lambda_s\pi{\rho^2}}/{N}$, and the probability mass function is given by
\begin{align}
\Pr\left(l=\kappa\right)=\frac{\left({\lambda_s{\pi}\rho^2}/{N}\right)^{\kappa}}{\kappa!}\exp\left(-{\lambda_s\pi\rho^2}/{N}\right),~\kappa=0,1,\cdots
\end{align}
The probability that no SN is inside a sector is thus
\begin{align}
p=\Pr\left(l=0\right)=\exp\left(-{\lambda_s\pi\rho^2}/{N}\right).\label{p}
\end{align}
Therefore, the active probability of a sector is the probability that at least one SN is inside this sector, which is given by
\begin{align}
q=1-p=1-\exp\left(-{\lambda_s\pi\rho^2}/{N}\right).
\end{align}
%, i.e.,
%\begin{align}
%\eta_n^M=\Pr\left[\text{SN$_0$ receives power from PB$_i$ with $G_M$}\mid|X_i|\leq{\rho}\right]\nonumber
%\end{align}
%and
%\begin{align}
%\eta_f^M=\Pr\left[\text{SN$_0$ receives power from PB$_i$ with $G_M$}\mid|X_i|>{\rho}\right]\nonumber.
%\end{align}

Denote $\eta_n^M$ and $\eta_f^M$ as the conditional probabilities that SN$_0$ receives energy from PB$_i$ with antenna gain $G_M$ given PB$_i$ is a near PB and a far PB, respectively.
Based on $p$, $q$, Conditions 1 and 2, we derive $\eta_n^M$ and $\eta_f^M$ as follows.
\subsubsection{Near PBs}
If $\|X_i\|\leq{\rho}$, PB$_i$ radiates energy in at least the direction towards SN$_0$ ($M\geq{1}$). Condition 2 is thus satisfied.
Given PB$_i$ is a near PB, the conditional probability that PB$_i$ radiates with gain $G_M$ is
\begin{align}
\omega_n^M=\binom{N-1}{M-1}p^{N-M}q^{M-1},
\end{align}
which is the probability that the rest $M-1$ out of $N-1$ sectors of PB$_i$ have SNs.
%For each $M$, there is a radius $\hat{\rho}$ that maximizes $\omega_n^M$. It can be shown that $\hat{\rho}$ increases as the increased $M$. In other words, the number of beams of the near PB increases as the increase $\rho$.
Given PB$_i$ is a near PB that radiates with gain $G_M$, the conditional probability that SN$_0$ receives energy from PB$_i$ is
\begin{align}
\varphi_n^M=1\label{probnear}.
\end{align}
Since $\eta_n^M=\varphi_n^M\omega_n^M$, we obtain the following lemma.
\begin{lemma}
Given PB$_i$ is a near PB, the conditional probability that SN$_0$ receives energy from PB$_i$ with gain $G_M$ is
\begin{align}
\eta_{n}^{M}
&=\binom{N-1}{M-1}p^{N-M}q^{M-1}\label{pp1},~\text{for}~M=1,\cdots,N.
\end{align}
\end{lemma}

\subsubsection{Far PBs}
If $\|X_i\|>{\rho}$, PB$_i$ may not radiate energy towards SN$_0$  ($M=0,\cdots,N$). SN$_0$ receives energy PB$_i$ with  $G_M$ if both Conditions 1 and 2 are satisfied.
% i.e.,
%\begin{align}
%&\mathbbold{1}\left(\text{SN$_0$ receives $G_M$ from PB$_i$}||X_i|>{\rho}\right)\nonumber\\
%&=\mathbbold{1}\left(\text{$M$ out of $N$ sectors of PB$_i$ have SNs}\right)\nonumber\\
%&~~~\times\mathbbold{1}\left(\text{SN$_0$ is within a direction of radiation}\right)\label{event2}.
%\end{align}
%If no SN is inside the power region of PB$_i$, PB$_i$ equally radiates energy in all directions with gain of $1$.
Given PB$_i$ is a far PB, the conditional probability that PB$_i$ radiates with gain $G_M$ is
\begin{subnumcases}{\omega_f^M=\nonumber}
p^{N},~~~~~~~~~~~~~~~~\text{for}~M=0 \\
\binom{N}{M}p^{N-M}q^{M},~\text{for}~M=1,\cdots,N.
\end{subnumcases}
%For each $M$, there is a radius $\hat{\rho}$ that maximizes $\omega_f^M$. It can be shown that $\hat{\rho}$ increases as the increased $M$. In other words, the number of beams of the far PB decreases and then increases as the increase $\rho$.
Given PB$_i$ is a far PB that radiates with gain $G_M$, the conditional probability that SN$_0$ receives energy from PB$_i$  is
\begin{subnumcases}{\varphi_f^M=\nonumber}
\label{probfar1} 1,~~~\text{for}~M=0 \\
\label{probfar2}\frac{M}{N},~\text{for}~M=1,\cdots,N.
\end{subnumcases}
Since $\eta_f^M=\varphi_f^M\omega_f^M$, we obtain the following lemma.
%It equals the joint probability of the event that $M$ out of $N$ sectors of PB$_i$ have SNs and the event that SN$_0$ is in a radiation direction of PB$_i$.
\begin{lemma}
Given PB$_i$ is a far PB, the conditional probability that SN$_0$ receives energy from PB$_i$ with gain $G_M$ is
\begin{subnumcases}{\eta_{f}^{M}=\nonumber}
\label{pp22}p^{N},~~~~~~~~~~~~~~~~~~~~~\text{for}~M=0 \\
\label{pp21}\binom{N-1}{M-1}p^{N-M}q^{M},~\text{for}~M=1,\cdots,N.
\end{subnumcases}
\end{lemma}
%We note that $\eta_{n}^{M}$ and $\eta_{f}^{M}$ are not related to the PB locations.

\subsection{Characterization of Received Power via Laplace Transform}
In this subsection, we derive the Laplace transform of the distribution of $P_s$ to characterize the received power at SN$_0$.
%The traditional method of derivation is similar to Section 3.2.2 in \cite{martinbook}. In the following discussions, we adopt another method by utilizing the thinning processes, which is shown to give the same results as the traditional method.
%Since the gains of the radiation energy of the PBs varies and depends on the distance from SN$_0$, the PB network can be regarded as a heterogeneous network.

Define $\Phi_{p}^M$ as the set of PBs with gain $G_M$ and $\Phi_{p}^{'}$ as the set of PBs that radiate energy towards SN$_0$.
The set of near PBs within ${b}(o,\rho)$ that radiate energy with gain $G_M$ towards SN$_0$ is
\begin{align}
\Phi_{p,n}^M=\Phi_{p}^M\bigcap\Phi_{p}^{'}\bigcap{b}(o,\rho),~\text{for}~M=1,\cdots,N,
\end{align}
which is obtained through the independent thinning \cite{stochastic} of near PBs with new density $\lambda_p\eta_{n}^{M}$, where $\eta_{n}^{M}$ is given in Lemma 1. The near PBs  can be regarded as a heterogeneous network consisting of a group of homogeneous networks each with antenna gain $G_M$ and density $\lambda_p\eta_{n}^{M}$.
Similarly, the set of far PBs  within $\overline{{b}(o,\rho)}$ that radiate energy with gain $G_M$ towards SN$_0$ is
\begin{align}
\Phi_{p,f}^M=\Phi_{p}^M\bigcap\Phi_{p}^{'}\bigcap{\overline{b(o,\rho)}},~\text{for}~M=0,\cdots,N,
\end{align}
which by the independent thinning of far PBs with new density  $\lambda_p\eta_{f}^{M}$, where $\eta_{f}^{M}$ is given in Lemma 2. The far PBs that radiate power towards SN$_0$ can be regarded as another heterogeneous network consisting of a group of homogeneous networks each with gain $G_M$ and density $\lambda_p\eta_{f}^{M}$. Note that SN$_0$ receives zero power from the far PBs that does not radiate energy towards SN$_0$. In the following, we derive the Laplace transform of the received power distribution in each homogeneous network, and then derive that of the aggregate received power from all PBs.

%Denote $P_{s,n}^{M}$ as the aggregate received power from the near PBs with gain $G_M$, and $P_{s,f}^{M}$
%as the aggregate received power from the far PBs with gain $G_M$.
%We rewrite the aggregate received power at SN$_0$ from all near PBs and all far PBs, respectively, as
%\begin{align}
%P_{s,n}=\sum\limits_{M=1}^{N}P_{s,n}^{M}=\sum\limits_{M=1}^{N}P_p\sigma\sum\limits_{X_i\in\Phi_{p,n}^M}{G}_M\left[\max\left(|X_i|,1\right)\right]^{-\alpha}\nonumber
%\end{align}
%and
%\begin{align}
%P_{s,f}=\sum\limits_{M=0}^{N}P_{s,f}^{M}=\sum\limits_{M=0}^{N}P_p\sigma\sum\limits_{X_i\in\Phi_{p,f}^M}{G}_M\left[\max\left(|X_i|,1\right)\right]^{-\alpha}.\nonumber
%\end{align}
We rewrite the aggregate received power at SN$_0$ from all the near PBs and far PBs in \eqref{dirins} as
\begin{align}
P_s=P_{s,n}+P_{s,f}=\sum\limits_{M=1}^{N}P_{s,n}^{M}+\sum\limits_{M=0}^{N}P_{s,f}^{M},
\end{align}
where
\begin{align}
&P_{s,n}^{M}=P_p\sigma\sum\limits_{X_i\in\Phi_{p,n}^M}{G}_M\left[\max\left(\|X_i\|,1\right)\right]^{-\alpha}\label{Psnnn}
\end{align}
is the aggregate received power from the near PBs with gain $G_M$ and
\begin{align}
&P_{s,f}^{M}=P_p\sigma\sum\limits_{X_i\in\Phi_{p,f}^M}{G}_M\left[\max\left(\|X_i\|,1\right)\right]^{-\alpha}\label{Psfff}
\end{align}
is the aggregate received power from the far PBs with gain $G_M$.
Since we adopt the non-singular path loss function $\left[\max\left(\|X_i\|,1\right)\right]^{-\alpha}$, our analysis  involves two cases: $0<\rho\leq1$ and $1<\rho<\infty$. Define $\gamma(s,x)=\int_0^{x}t^{s-1}e^{-t}dt$ as the lower incomplete gamma function.
The Laplace transforms of the distributions of $P_{s,n}^{M}$ and $P_{s,f}^{M}$ are given as follows.
\begin{lemma}
The Laplace transform of the distribution of aggregate received power  at the typical SN$_0$ from the near PBs with gain $G_M$ is
\begin{subnumcases}{\mathcal{L}_{P_{s,n}^{M}}(s)=\nonumber\\}
\label{1}\mathcal{L}_{P_{s,n}^{M}(1)}(s), ~\text{for}~0<\rho\leq1\\
\label{2}\mathcal{L}_{P_{s,n}^{M}(2)}(s), ~\text{for}~1<\rho<\infty,
\end{subnumcases}
where
\begin{align}
\mathcal{L}_{P_{s,n}^{M}(1)}(s)=\exp\Bigg\{-\lambda_p\pi{\eta}_{n}^{M}\left[\rho^2-\rho^2\exp\left({-s{P}_p\sigma{G}_M}\right)\right]\Bigg\}\label{LL1}
\end{align}
and
\begin{align}
\mathcal{L}_{P_{s,n}^{M}(2)}(s)
&=\exp\Bigg\{-\lambda_p\pi{\eta}_{n}^{M}\bigg\{\rho^2-\rho^2\exp\left({-s{P}_p\sigma{G}_M{\rho}^{-\alpha}}\right)\nonumber\\
&~~~~~~+\left(s{P}_p\sigma{G}_M\right)^{\frac{2}{\alpha}}\left[\gamma\left(1-\frac{2}{\alpha},s{P}_p\sigma{G}_M\right)-\gamma\left(1-\frac{2}{\alpha},s{P}_p\sigma{G}_M\rho^{-\alpha}\right)\right]\bigg\}\Bigg\}.\label{LL2}
\end{align}
\end{lemma}
%{\bf{Proof}}:
\begin{IEEEproof}
See Appendix A.
\end{IEEEproof}
\begin{lemma}
The Laplace transform of the distribution of  aggregate received power at the typical SN$_0$ from the far PBs with gain $G_M$ is
\begin{subnumcases}{\mathcal{L}_{P_{s,f}^{M}}(s)=\nonumber\\}
\label{1}\mathcal{L}_{P_{s,f}^{M}(1)}(s),~\text{for}~0<\rho\leq1\label{L1}\\
\label{2}\mathcal{L}_{P_{s,f}^{M}(2)}(s),~\text{for}~1<\rho<\infty,\label{L2}
\end{subnumcases}
where
\begin{align}
\mathcal{L}_{P_{s,f}^{M}(1)}(s)=\exp\Bigg\{\lambda_p\pi{\eta}_{f}^{M}\bigg\{\rho^2-\rho^2\exp\left({-s{P}_p\sigma{G}_M}\right)
-\left({s{P}_p\sigma{G}_M}\right)^{\frac{2}{\alpha}}\gamma\left(1-\frac{2}{\alpha},s{P}_p\sigma{G}_M\right)\bigg\}\Bigg\}\label{LL3}
\end{align}
and
\begin{align}
\mathcal{L}_{P_{s,f}^{M}(2)}(s)
=&\exp\Bigg\{\lambda_p\pi{\eta}_{f}^{M}\bigg\{\rho^2-\rho^2\exp\left({-s{P}_p\sigma{G}_M{\rho}^{-\alpha}}\right)\nonumber\\
&~~~~~~~-\left({s{P}_p\sigma{G}_M}\right)^{\frac{2}{\alpha}}\gamma\left(1-\frac{2}{\alpha},s{P}_p\sigma{G}_M{\rho}^{-\alpha}\right)\bigg\}\Bigg\}\label{LL4}.
\end{align}
\end{lemma}
\begin{IEEEproof}
See Appendix B.
\end{IEEEproof}
Based on Lemma 3 and Lemma 4, we obtain the Laplace transform of the distribution of $P_s$ in the following proposition.
\begin{proposition}
The Laplace transform of the distribution of aggregate received power at the typical SN$_0$ from all PBs under AD-WPT is given by
\begin{subnumcases}{\mathcal{L}_{P_s}(s)=\nonumber\\}
\label{Lap1}\prod\limits_{M=1}^{N}\mathcal{L}_{P_{s,n}^{M}(1)}\left(s\right)\prod\limits_{M=0}^{N}\mathcal{L}_{P_{s,f}^{M}(1)}\left(s\right),~\text{for}~0<\rho\leq1\label{Ldir1}\\
\label{Lap2}\prod\limits_{M=1}^{N}\mathcal{L}_{P_{s,n}^{M}(2)}\left(s\right)\prod\limits_{M=0}^{N}\mathcal{L}_{P_{s,f}^{M}(2)}\left(s\right),~\text{for}~1<\rho<\infty\label{Ldir2}.
\end{subnumcases}
As a special case of $N=1$, the Laplace transform of the distribution of aggregate received power at SN$_0$ from all PBs in omnidirectional WPT is given by
\begin{align}
\mathcal{L}_{P_s^{omni}}(s)
&=\exp\Bigg\{-\lambda_p\pi\left(s{P}_p\sigma\right)^{\frac{2}{\alpha}}\gamma\left(1-\frac{2}{\alpha},s{P}_p\sigma\right)\Bigg\}.
%\label{Liso}
\end{align}
\end{proposition}
\begin{IEEEproof}
\begin{align}
\mathcal{L}_{P_s}(s)&=\mathbb{E}\left[\exp\left(-sP_s\right)\right]=\mathbb{E}\left[\exp\left(-s\left(P_{s,n}+P_{s,f}\right)\right)\right]=\mathbb{E}\left[\exp\left(-sP_{s,n}\right)\right]\mathbb{E}\left[\exp\left(-sP_{s,f}\right)\right]\nonumber\\
&=\mathbb{E}\left[\exp\left(-s\sum\limits_{M=1}^{N}P_{s,n}^{M}\right)\right]\mathbb{E}\left[\exp\left(-s\sum\limits_{M=0}^{N}P_{s,f}^{M}\right)\right]
%=\prod\limits_{M=1}^{N}\mathbb{E}\left[\exp\left(-sP_{s,n}^{M}\right)\right]\prod\limits_{M=0}^{N}\mathbb{E}\left[\exp\left(-sP_{s,f}^{M}\right)\right]\nonumber\\
=\prod\limits_{M=1}^{N}\mathcal{L}_{P_{s,n}^{M}}(s)\prod\limits_{M=0}^{N}\mathcal{L}_{P_{s,f}^{M}}(s).\label{LPsnf}
\end{align}
Substituting $\mathcal{L}_{P_{s,n}^{M}}(s)$ in Lemma 3 and $\mathcal{L}_{P_{s,f}^{M}}(s)$ in Lemma 4  into \eqref{LPsnf}, we obtain the Laplace transform given in Proposition 1. It is noted that $\mathcal{L}_{P_s}(s)$ is continuous at ${\rho}=1$.
\end{IEEEproof}

\subsection{Mean and Variance of Received Power}
In this subsection, we derive the closed-form mean and variance of the received power at SN$_0$ by taking the derivative of the Laplace transform in Proposition 1. These results will be useful in the approximation of the CCDF of received power in the next subsection and the average power maximization in Section IV.

The average received power at SN$_0$ is given by
\begin{align}
\mathbb{E}[P_s]=-\frac{d}{ds}\left[\log\left(\mathcal{L}_{P_{s}}(s)\right)\right]|_{s=0}.
\end{align}
This is the expectation of the aggregate received power at SN$_0$ from all PBs by taking over all possible location realizations of the PBs in spatial domain.
%, which is equivalent to the average received power over all SNs in the network.
By further derivations, the results are summarized in the following proposition.
\begin{proposition}
At the typical SN$_0$, the average received power in AD-WPT is given by
\!\begin{subnumcases}{\mathbb{E}\left[P_s\right]=\nonumber\\}
\label{E1}{P_p\lambda_p}{\sigma\pi}\left[\frac{\rho^2\left(p-p^N\right)}{1-p}+\frac{\alpha}{\alpha-2}\right],~~~~~~~~~~~~~\text{for}~0<\rho\leq1\\
\label{E2}{P_p\lambda_p}{\sigma\pi}\left[\frac{\left(\alpha-2{\rho}^{2-\alpha}\right)\left(1-p^N\right)}{\left(\alpha-2\right)\left(1-p\right)}+\frac{2{\rho}^{2-\alpha}}{\alpha-2}\right],~\text{for}~1<\rho<\infty,
\end{subnumcases}
where $p$ is given in \eqref{p} and $\mathbb{E}(P_s)$ is continuous at ${\rho}=1$.
As a special case of $N=1$, the average received power at SN$_0$ in omnidirectional WPT is given by
\begin{align}
\mathbb{E}[P_s^{omni}]=\frac{{P_p\lambda_p}{\sigma\pi}\alpha}{\alpha-2}\label{Eiso}.
\end{align}
\end{proposition}
\begin{IEEEproof}
See Appendix C. With $N=1$, both \eqref{E1} and \eqref{E2} equal \eqref{Eiso} for all $\rho$.
\end{IEEEproof}
In Proposition 2, for any given $\rho$, $\mathbb{E}[P_s]$ is increasing with the increased $P_p$, $\lambda_p$ and $N$ and is decreasing with the increased $\lambda_s$. Moreover, for any given set of $\{P_p,\lambda_p,\lambda_s,N\}$, $\mathbb{E}[P_s]$ is unimodal in $\rho$, i.e., there exists a unique $\rho^*$ that maximizes $\mathbb{E}[P_s]$, where $\mathbb{E}[P_s]$ is monotonically increasing for $\rho\leq\rho^*$ and is monotonically decreasing for $\rho\geq\rho^*$. In Section IV, we will analyze the optimal $\rho^*$ that maximizes $\mathbb{E}[P_s]$.

The comparison of the average received power between AD-WPT and omnidirectional WPT is given in the following corollaries.
%\begin{remark}
\begin{corollary}
For $0<\rho<\infty$, it follows that $\mathbb{E}[P_s]>\mathbb{E}[P_s^{omni}]$. For $\rho\rightarrow0$ and $\rho\rightarrow\infty$, we have $\mathbb{E}[P_s]\rightarrow\mathbb{E}[P_s^{omni}]$.
\end{corollary}
\begin{IEEEproof}
See Appendix D.
\end{IEEEproof}
From Corollary 1, we see that the average received power at SN$_0$ from all PBs in AD-WPT is higher than that in omnidirectional WPT.
%Next, we further compare the the average received power between AD-WPT and omnidirectional WPT for the near PBs ($|X_i|<\rho$) and far PBs ($|X_i|\geq\rho$), respectively.
Next, we further discuss how the near PBs ($\|X_i\|\leq\rho$) and far PBs ($\|X_i\|>\rho$) influence the average received power.
%We denote the omnidirectional near PBs and far PBs as the omnidirectional radiators with distance $|X_i|<\rho$ and $|X_i|\geq\rho$ from SN$_0$, respectively.
For comparison, we denote the aggregate received power from the PBs with $\|X_i\|\leq\rho$ and $\|X_i\|>\rho$ in omnidirectional  WPT by $P_{s,n}^{omni}$ and $P_{s,f}^{omni}$, respectively.
\begin{corollary}
The ratio of the average received power at SN$_0$ from the near PBs in AD-WPT and in omnidirectional WPT  is given by
\begin{align}
\frac{\mathbb{E}[P_{s,n}]}{\mathbb{E}[P_{s,n}^{omni}]}=\frac{1-p^N}{1-p}\geq1 \label{EPsnratio}
\end{align}
The ratio of the average received power at SN$_0$ from the far PBs in AD-WPT and in omnidirectional WPT is given by
\begin{align}
\frac{\mathbb{E}[P_{s,f}]}{\mathbb{E}[P_{s,f}^{omni}]}=1 \label{EPsfratio}.
\end{align}
%$\mathbb{E}(P_{s,n})$ is unimodal in $\rho$ and $\mathbb{E}(P_{s,f})$ is a decreasing function of $\rho$.
\end{corollary}
\begin{IEEEproof}
The proof is similar to that of Proposition 2 and thus omitted.
\end{IEEEproof}
%From Corollary 2, it is seen that the average received power from the near PBs with AD-WPT is higher than that with omnidirectional WPT. However, the far PBs with AD-WPT and omnidirectional WPT give equal contributions to the average received power at SN$_0$ %\subsection{Variance of the Received Power}
In Corollary 2, we see that the average received power at SN$_0$ from the near PBs and far PBs in AD-WPT is greater than and equal to that in omnidirectional WPT as shown in \eqref{EPsnratio} and \eqref{EPsfratio}, respectively.
The improvement of the average received power from all PBs at SN$_0$ is thus due to the adaptive energy beamforming of the near PBs.

%\subsection{Variance of the Received Power}
By taking the second derivative of the Laplace transform in Proposition 1, we obtain the variance of the received power
at SN$_0$, i.e.,
\begin{align}
\mathbb{V}[P_s]=\frac{d^2}{ds^2}\left[\log\left(\mathcal{L}_{P_{s}}(s)\right)\right]|_{s=0}.
\end{align}
By further derivations, we summarize the results in the following proposition.
\begin{proposition}
At the typical SN$_0$, the variance of the received power in AD-WPT is given by
\begin{subnumcases}{\mathbb{V}\left[P_s\right]=\nonumber\\}
\label{v1}{{\lambda_p}P_p^2}\sigma^2{\pi}\Bigg\{\left[\frac{\alpha}{\alpha-1}-\rho^2\right]p^N+\left[\left(q^{-1}-1\right)\rho^2+\frac{\alpha}{\alpha-1}\right]\nonumber\\
~~~~~~~~~~~~~~\times\sum\limits_{M=1}^N\left(\frac{N}{M}\right)^2\binom{N-1}{M-1}p^{N-M}q^{M}\Bigg\},~\text{for}~0<\rho\leq1\label{V1}\\
\label{v2}{{\lambda_p}P_p^2}\sigma^2{\pi}\Bigg\{\frac{\rho^{2-2\alpha}}{\alpha-1}p^N+\left(\frac{\alpha-\rho^{2-2\alpha}}{\alpha-1}q^{-1}+\frac{\rho^{2-2\alpha}}{\alpha-1}\right)\nonumber\\
~~~~~~~~~~~~~~\times\sum\limits_{M=1}^N\left(\frac{N}{M}\right)^2\binom{N-1}{M-1}p^{N-M}q^{M}\Bigg\},\text{for}~1<\rho<\infty,\label{V2}
\end{subnumcases}
 where $\mathbb{V}(P_s)$ is continuous at ${\rho}=1$.
As a special case of $N=1$, the variance of the received power in omnidirectional WPT is
\begin{align}
\mathbb{V}[P_s^{omni}]=\frac{{{\lambda_p}P_p^2}\sigma^2{\pi}\alpha}{\alpha-1}\label{Viso}.
\end{align}
\end{proposition}
\begin{IEEEproof}
See Appendix E. With $N=1$, both \eqref{V1} and \eqref{V2} equal \eqref{Viso} for all $\rho$.
\end{IEEEproof}
%In Proposition 3, $\mathbb{V}[P_s]$ is unimodal in $\rho$, i.e., $\mathbb{V}[P_s]$ is first  increasing and then decreasing with the increased $\rho$. %Given any $\rho$, $\mathbb{V}[P_s]$ is increasing with $P_p$, $\lambda_p$ and $N$, and it is decreasing with $\lambda_s$.
%We also compare
%Similar to Corollary 1, it can be proved that $\mathbb{V}[P_s]>\mathbb{V}[P_s^{omni}]$ for $0<\rho<\infty$.
%Though AD-WPT improves the average received power compared with omnidirectional WPT, but it also causes more significant spatial  power fluctuation.

In Proposition 3, $\mathbb{V}[P_s]$ is unimodal in $\rho$, i.e., $\mathbb{V}[P_s]$ is first  increasing and then decreasing with the increased $\rho$. Given any $\rho$, $\mathbb{V}[P_s]$ is increasing with $P_p$, $\lambda_p$ and $N$, and it is decreasing with $\lambda_s$. We also compare the variance of the received power at SN$_0$ between AD-WPT and omnidirectional WPT in the following corollary.
%\eqref{V1} equals \eqref{V2} for $\rho=1$.
\begin{corollary}
For $\rho\rightarrow0$ and $\rho\rightarrow\infty$, we have $\mathbb{V}[P_s]\rightarrow\mathbb{V}[P_s^{omni}]$. For $0<\rho<\infty$, it follows that $\mathbb{V}[P_s]>\mathbb{V}[P_s^{omni}]$.
\end{corollary}
\begin{IEEEproof}
The proof is similar to that of Corollary 1 and thus omitted.
\end{IEEEproof}
From Corollary 3, we see that variance of received power in AD-WPT  is higher than that in omnidirectional WPT. Though AD-WPT improves the average received power compared with omnidirectional WPT, it also causes more significant spatial  power fluctuation.

\subsection{Characterization of Received Power via CCDF}
% add references
\begin{figure}[t!]
    \begin{center}
        \includegraphics[width=0.65\columnwidth]{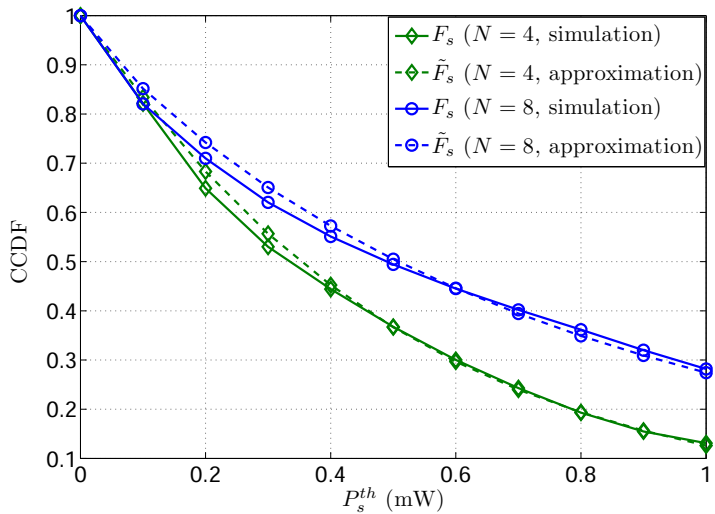}
        \caption{CCDF of the received power at SN$_0$ ($P_p=5$ W, $\rho=2$, $\lambda_p=0.1$, $\lambda_s=0.2$, $\nu=0.1$ m and $\alpha=3$).}
    \end{center}
\end{figure}
In this subsection, we analyze the CCDF $F_s$ of the received power at the typical SN$_0$, which is the probability that $P_s$  takes on a value greater than or equal to the threshold $P_{s}^{th}$, i.e.,
\begin{align}
F_s&=\Pr\left(P_s\geq{P_{s}^{th}}\right)=\int_{P_{s}^{th}}^{\infty}f(P_s)dP_s,\label{tau}
\end{align}
where $f(P_s)$ is the probability density function (PDF) of $P_s$ and can be calculated from the inverse Laplace transform of $\mathcal{L}_{P_s}(s)$ in Proposition 1, i.e.,
\begin{align}
f(P_s)=\mathcal{L}_{P_s}^{-1}(s).
\end{align}
In most cases, the direct derivation of the PDF from the inverse Laplace transform is very challenging, if not possible, especially for the non-singular path loss model. In our problem, the new parameter $\rho$ adds more difficulty to the derivation.
Even for omnidirectional transmission, the closed-form PDF only exists for special choices of parameters, e.g., $\alpha=4$ and singular path loss model \cite{martinbook}.
%In our problem, since $\mathcal{L}_{P_s}(s)$ is a summation of several groups of incomplete Gamma function, the analytical derivation of CCDF from the inverse Laplace transform is not analytically tractable.
% add reference...maybe robert heath or book...
%Due to the difficulty in deriving a tractable analytical $\tau_s$,
Studies have been shown that Gamma distribution gives a good fit to the power distribution for the homogeneous PPP \cite{ganti} and heterogeneous PPP  \cite{gamma} with non-singular path loss.
In this work, we approximate the received power distribution by Gamma distribution with second-order moment matching method, i.e., by matching the mean and variance of the two distributions, where the mean and variance of $P_s$ are given in Proposition 2 and Proposition 3, respectively.
%Before deriving the Gamma approximation, we firstly analyze the variance of $P_s$.
Denote the Gamma function by $\Gamma(k)=\int_{0}^{\infty}x^{k-1}e^{-t}dt$. The approximation of the CCDF of $P_s$ is given in the following proposition.
%The PDF and $\tau$ of Gamma distribution is
%\begin{align}
%f_{\text{Gamma}}(P_{s}^{th})=\frac{{P_s}^{k-1}e^{-\frac{P_s}{\theta}}}{\theta^{k}\Gamma(k)}, ~~~ x>0
%\end{align}
%and
\begin{proposition}
At the typical SN$_0$, the approximated CCDF of the received power using Gamma distribution with second-order matching is expressed as
\begin{align}
\tilde{F}_s=1-\frac{\gamma\left(k,\frac{P_s^{th}}{\theta}\right)}{\Gamma\left(k\right)}, \label{tildeFs}
\end{align}
where $k=\frac{\left[\mathbb{E}\left[P_{s}\right]\right]^2}{\mathbb{V}\left[P_{s}\right]}$ and $\theta=\frac{\mathbb{V}\left[P_{s}\right]}{\mathbb{E}\left[P_{s}\right]}$ are the shape parameter and scale parameter of the Gamma distribution, respectively.
%\label{k}
%and
%\begin{align}
%\label{theta}
%\end{align}
%are the shape parameter and scale parameter of the Gamma distribution.
\end{proposition}

Fig. 2 shows a good match between the simulation results of $F_s$ and its approximation $\tilde{F}_s$.
%As a special case of $N=1$, the CDF of the received power for omnidirectional WPT $\text{CDF}_s^{omni}$ can be derived by substituting $P_s^{omni}$ in \eqref{Psiso} into \eqref{tau}. Its approximation $\tilde{\text{CDF}}_s^{omni}$ is obtained by substituting $\mathbb{E}\left(P_{s}^{omni}\right)$ in \eqref{Eiso} and $\mathbb{V}\left(P_{s}^{omni}\right)$ in \eqref{Viso} into Proposition 4.
%We denote  the complementary cumulative distribution function (CCDF) of $P_s$ as $F_s$ and its approximation as $\tilde{F}_s$, where $F_s=1-\text{CDF}_s$ and $\tilde{F}_s=1-\tilde{\text{CDF}_s}$.
In some scenarios, an SN is active if the received power is beyond the constant operational power threshold $P_s^{th}$. Then, the CCDF of $P_s$ can be regarded as the active probability of the SNs. It can be proved that $\tilde{F}_s$ in \eqref{tildeFs} is increasing in $\mathbb{E}[P_s]$ and decreasing in $\mathbb{V}[P_s]$.
%As discussed in the previous subsections, both $\mathbb{E}(P_s)$ and $\mathbb{V}(P_s)$ first increase and then decrease with the increased of $\rho$.
The increased average received power and power fluctuation may improve or reduce the sensor active probability, respectively. As shown in Proposition 2 and 3, both $\mathbb{E}[P_s]$ and $\mathbb{V}[P_s]$ first increase and then decrease with the increased $\rho$.
%The increase of average received power $\mathbb{E}(P_s)$ also leads to higher power fluctuation $\mathbb{V}(P_s)$.
We will further discuss the above tradeoff and derive the optimal $\rho^*$ that maximizes the sensor active probability in Section V.
For omnidirectional WPT, the CCDF of received power $F_s^{omni}$ and its approximation $\tilde{F}_s^{omni}$ can be obtained by substituting \eqref{Psiso} into \eqref{tau} and by substituting \eqref{Eiso} and \eqref{Viso} into \eqref{tildeFs}, respectively.

\begin{corollary}
$\tilde{F}_s$ increases with the increased $P_p$ and/or $\lambda_p$.
\end{corollary}
\begin{IEEEproof}
It can be proved that $\theta$ and $k$ are linear increasing with $P_p$ and $\lambda_p$, respectively. As $P_p$ increases, $\theta$ increases and $k$ remains a constant. Since $\tilde{F}_s$ is an increasing function of $\theta$, it also increases with the increased of $P_p$. Similarly, as $\lambda_p$ increases, $k$ increases and $\theta$ remains a constant. Since $\tilde{F}_s$ is an increasing functions of $k$, it also increases with the increased $\lambda_p$.
\end{IEEEproof}
Corollary 4 shows that increasing the PB density or power improves the sensor active probability.

\section{Maximization of Average Received Power in AD-WPT}
\begin{figure}[t!]
    \begin{center}
        \includegraphics[width=0.63\columnwidth]{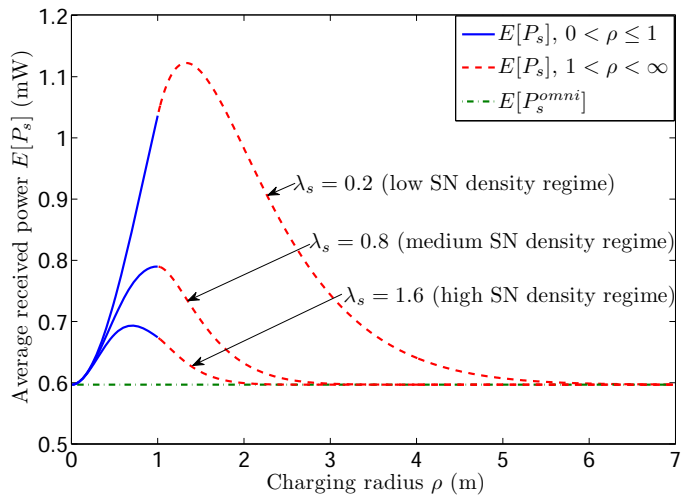}
        \caption{Average received power versus charging radius ($P_p=10$ W, $\nu=0.1$ m, $\alpha=3$, $\lambda_p=0.1$ and $N=4$).}
    \end{center}
\end{figure}

%Some WSN are task flexible, where each SN is assigned with flexible amounts of sensing tasks depending on the received energy, i.e., the SNs with high received power may handle more tasks for the benefit of the whole network.
%For example, the high power SNs in a hierarchical network may work as cluster-heads \cite{cluster} that collect data from the low power SNs and coordinate sensing tasks among the SNs \cite{IFsensor}.
%To achieve high energy efficient AD-WPT in a flexible-task WSN, it is important to design the optimal charging radius that maximizes the average received energy over all SNs in the network. In this section, we design the optimal $\rho^*$  for maximizing $\mathbb{E}(P_s)$ in Proposition 2.

In flexible-task WSN applications, each SN is assigned with flexible sensing tasks depending on the received energy, i.e., the SNs with high received power may handle more tasks than the low-power SNs for the benefit of the whole network.
For example, the high-power SNs in a hierarchical network may work as cluster-heads \cite{IFsensor} that collect data from the low-power SNs and coordinate sensing tasks among the SNs. The low-power SNs can also offload part of the computational processing tasks to the high-power SNs which have abundant resource  \cite{sigmag}.
To achieve energy efficient AD-WPT in flexible-task WSN, it is important to maximize the total received power over all SNs, which is equivalent to maximizing the average received power at the typical SN$_0$.
In this section, we design the optimal charging radius $\rho^*$  for maximizing $\mathbb{E}[P_s]$ in Proposition 2.

In Fig. 3, we plot the average received power  $\mathbb{E}[P_s]$ in AD-WPT, which outperforms the average received power  $\mathbb{E}[P_s^{omni}]$ in omnidirectional WPT for all $\rho>0$. Moreover, $\mathbb{E}[P_s]$ first  increases and then decreases with the increased $\rho$ and there exists an optimal $\rho^*$ that maximizes $\mathbb{E}[P_s]$. These results  match well with  Proposition 2 and Corollary 1 and are explained as follows.
The received power at SN$_0$ from PB$_i$ depends on whether PB$_i$ radiates power towards SN$_0$ and the intensity of the radiated energy, which can be viewed as the power opportunity and power intensity of the PB, respectively. Both the power opportunity and power intensity are related to the number of beams of PB$_i$ and the distance $\|X_i\|$ between PB$_i$ and SN$_0$. We discuss the average received power at SN$_0$ from the near PBs ($\|X_i\|\leq\rho$) and far PBs ($\|X_i\|>\rho$) as follows.
\begin{itemize}
\item \textit{Near PBs}: The average received power from the near PBs in AD-WPT is higher than that in omnidirectional WPT (see \eqref{EPsnratio} in Corollary 2).
    When the number of beams from  a near PB decreases, the power intensity of this near PB increases (see \eqref{Gmn}), and the power opportunity that the near PB radiates power towards SN$_0$ is with probability $1$  since the near PB forms at least one beam towards SN$_0$.
   %When a near PB is directional (with small number of beams), the power intensity of this near PB is high and the power opportunity is the same as the omnidirectional PB.
    %Therefore, the average received power at SN$_0$ from the near PBs increases with the decreased number of beams.
    %$P_{s,n}$ increases as the decreased total bandwidth, the average received power from the near PBs increases and then decreases as the increased $\rho$.
    %From the average perspective, the improved power intensity of the near PBs in AD-WPT is the main reason for the improvement of average received power of near PBs compared with that in the omnidirectional WPT.
\item \textit{Far PBs}:  The average received power from the far PBs in AD-WPT is the same as that in omnidirectional WPT (see \eqref{EPsfratio} in Corollary 2). When the number of beams of a far PB decreases, the power intensity of this far increases (see \eqref{Gm0} and \eqref{Gmn}), but the power opportunity that SN$_0$ receives energy from this far PBs decreases (see \eqref{probfar1} and \eqref{probfar2}), and vice versa. From the average perspective, the effects of power intensity and power opportunity of the far PBs cancel with each other.
    %, which makes the average received power from the far PBs in AD-WPT equivalent to that in the omnidirectional WPT.
    %Thus, the average received power at SN$_0$ from the far PBs  is regardless of the number of beams.
\end{itemize}
%From the above discussions, $\mathbb{E}(P_s)$ outperforms $\mathbb{E}(P_s^{omni})$ mainly because of the high power intensity from the near PBs.
%As $\rho\rightarrow{0}$ or $\rho\rightarrow\infty$ (when no SN or all SNs are in the charging regions), the average received power in AD-WPT is equivalent to that in omnidirectional WPT since all PBs radiate power in $N$ directions. As $\rho$ increases from $0$ to $\infty$, more PBs are regarded as near PBs at SN$_0$ and the number of SNs in the charging regions increases. Since the number of beams of most PBs first decreases and then increases as the increase of $\rho$, the corresponding power intensity increases and then decreases. $\mathbb{E}(P_s)$ first increases with the increased $\rho$ due to the increased number of near PBs and increased power intensity. As $\rho$ further increases, $\mathbb{E}(P_s)$ increases and then decreases due to the tradeoff between the increased number of near PBs and decreased power intensity.
From the above discussions, $\mathbb{E}[P_s]$ outperforms $\mathbb{E}[P_s^{omni}]$ mainly because of the high power intensity from the near PBs.
%As $\rho$ increases from $0$ to $\infty$, the number of beams of most PBs decreases and then increases, and the power intensity increases and then decreases (see Section II-B).
We discuss $\mathbb{E}[P_s]$ as follows.
%As $\rho$ increases from $0$ to $\infty$, the number of beams of most PBs decreases from $N$ to $1$, and then increases from $1$, $2$, $\cdots$, to $N$. The corresponding power intensity of the PBs increases and then decreases, which therefore improves and reduces the average received power at SN$_0$.
\begin{itemize}
\item As $\rho\rightarrow{0}$, no SN is in the charging regions. $\mathbb{E}[P_s]$ is equivalent to $\mathbb{E}[P_s^{omni}]$ since all PBs radiate power in $N$ directions.
\item As $\rho$ increases, a small number of SNs are included in the charging regions and the PBs that are close to SN$_0$ become near PBs. When  most PBs concentrate the transmit power from $N$ beams into $1$ beam, the power intensity is greatly enhanced compared with omnidirectional WPT. $\mathbb{E}[P_s]$ increases with the increased $\rho$ due to the increased number of the near PBs and increased power intensity of the PBs.
\item As $\rho$ further increases, more sectors of the PBs are likely to be activated due to the increased number of SNs in the charging regions.
When the number of beams of most PBs increases from $1$, $2$, $\cdots$, to $N$, the power intensity for each beam decreases. There is a tradeoff between the further increased number of the near PBs and the decreased power intensity. $\mathbb{E}[P_s]$ thus increases and then decreases with the increased $\rho$.
\item As $\rho\rightarrow{\infty}$, all SNs are in the charging regions and AD-WPT is again equivalent to omnidirectional WPT.
\end{itemize}

%\subsection{Optimal Charging Radius that Maximizes Average Received Power}
%\subsection{Optimal Power Region Radius}
In the following, we study the optimal charging radius $\rho^*$ that maximizes $\mathbb{E}[P_s]$ in Proposition 2, i.e.,
\begin{eqnarray}
   \text{P1}:\ &&\mathbb{E}[P_s]^*=\max\limits_{0<\rho<{\infty}}~~{\mathbb{E}[P_s]}.
   %&&s.t.~~~~~0<\rho<{\infty}.
\end{eqnarray}
In Fig. 3, for omnidirectional WPT, $\mathbb{E}[P_s^{omni}]$ is regardless of $\lambda_s$ which matches with \eqref{Eiso}. It is because each PB radiates energy in all directions without catering to the locations or density of the SNs. For AD-WPT, $\mathbb{E}[P_s]$ decreases with the increased $\lambda_s$. With the increased number of SNs in the charging regions, PBs are more likely to radiate with more beams and less power intensity, which thus reduces the average received power at SN$_0$. In the following, we discuss the optimal $\rho^*$ for different $\lambda_s$ under AD-WPT.\\
\textit{Case 1: Low SN Density Regime}.
When the SN network density is low, e.g., $\lambda_s=0.2$ in Fig. 3, we have $\frac{\partial\mathbb{E}[P_s]}{\partial\rho}|_{\rho=1}>0$. Since \eqref{E2} is unimodal in $\rho$ and \eqref{E1} is an increasing function of $\rho$, the optimal charging radius $\rho^*\in(1,\infty)$ is the stationary point of \eqref{E2}.
Taking the first derivative of \eqref{E2} with respect to $\rho$, we have
\begin{align}
\frac{\partial\left[\mathbb{E}[P_s]|_{1<\rho<\infty}\right]}{\partial{\rho}}
&=2{P_p\lambda_p}{\pi}\sigma\left[\frac{{\rho}^{1-\alpha}\left(p-p^N\right)}{1-p}+\frac{\lambda_s\pi{p}^{N}{\rho}\left(\alpha-2{\rho}^{2-\alpha}\right)}{(\alpha-2)(1-p)}\right.\nonumber\\
&~~~~~~~~~~~~~~\left.-\frac{{\lambda_s\pi\rho}p\left(\alpha-2{\rho}^{2-\alpha}\right)\left(1-p^N\right)}{(1-p)^2(\alpha-2)N}\right].
\end{align}
%\begin{small}
%\begin{align}
%\frac{\partial\left[\mathbb{E}[P_s]|_{1<\rho<\infty}\right]}{\partial{\rho}}
%=2{P_p\lambda_p}{\pi}\sigma\left[\frac{{\rho}^{1-\alpha}\left(p-p^N\right)}{1-p}+\frac{\lambda_s\pi{p}^{N}{\rho}\left(\alpha-2{\rho}^{2-\alpha}\right)}{(\alpha-2)(1-p)}
%-\frac{{\lambda_s\pi\rho}p\left(\alpha-2{\rho}^{2-\alpha}\right)\left(1-p^N\right)}{(1-p)^2(\alpha-2)N}\right].\nonumber
%\end{align}
%\end{small}
The optimal charging radius  $\rho^*$ is the unique solution of $\frac{\partial\mathbb{E}[P_s]|_{\rho\geq1}}{\partial{\rho}}=0$. Though $\rho^*$ is not analytically tractable, we can search it numerically using one-dimensional searching method.\\
\textit{Case 2: Medium SN Density Regime}.
When the SN network has a medium density, e.g., $\lambda_s=0.8$ in Fig. 3, we have $\frac{\partial\mathbb{E}[P_s]}{\partial\rho}|_{\rho=1}=0$. In this case, \eqref{E1} is an increasing function of $\rho$ and \eqref{E2} is a decreasing function of $\rho$. The optimal charging region radius is at the point of $\rho^*=1$.\\
\textit{Case 3: High SN Density Regime}.
When the SN network has a high density, e.g., $\lambda_s=1.6$ in Fig. 3, we have $\frac{\partial\mathbb{E}[P_s]}{\partial\rho}|_{\rho=1}<0$. In this case, \eqref{E1} is unimodal in $\rho$ and \eqref{E2} is a decreasing function of $\rho$. The optimal charging radius $\rho^*\in(0,1)$ is the stationary point of \eqref{E1}.
Taking the first derivative of \eqref{E1}, we have
\begin{align}
&\frac{\partial\left[\mathbb{E}[P_s]|_{0<\rho\leq1}\right]}{\partial{\rho}}={2P_p\lambda_p}{\pi}\sigma\left[\frac{\lambda_s\pi\rho^3p(p^N-p)}{N(1-p)^2}
+\frac{\rho(p-p^N)+{\lambda_s\pi}\rho^3p\left(p^{N-1}-\frac{1}{N}\right)}{1-p}\right].
\end{align}
The optimal charging radius  $\rho^*$ is the unique solution to $\frac{\partial\left[\mathbb{E}[P_s]|_{0<\rho\leq1}\right]}{\partial{\rho}}=0$. Similar to Case 1, $\rho^*$ is not analytically tractable but can be searched numerically.

It can be proved that $\frac{\partial\left[\mathbb{E}[P_s]|_{0<\rho\leq1}\right]}{\partial{\rho}}$ and $\frac{\partial\left[\mathbb{E}[P_s]|_{1<\rho<\infty}\right]}{\partial{\rho}}$ are of the same sign at the point of $\rho=1$. The procedure of obtaining the optimal $\rho^*$ is summarized in Algorithm 1. More numerical results will be shown in Section VI-A.
\begin{algorithm}[h]
 \caption{Solving the optimal charging radius in P1:}
\begin{algorithmic}[1]
  \STATE Calculate $D_1(\rho)=\frac{\partial\left[\mathbb{E}[P_s]|_{0<\rho\leq1}\right]}{\partial{\rho}}$ and $D_2(\rho)=\frac{\partial\left[\mathbb{E}[P_s]|_{1<\rho<\infty}\right]}{\partial{\rho}}$
  \IF{either $D_1\left(\rho=1\right)<0$ or $D_2\left(\rho=1\right)<0$}
    \STATE $\rho^*$ is the solution to $D_1(\rho)=0$
  \ELSIF{either $D_1\left(\rho=1\right)=0$ or $D_2\left(\rho=1\right)=0$}
    \STATE $\rho^*=1$
  \ELSIF{either $D_1\left(\rho=1\right)>0$ or $D_2\left(\rho=1\right)>0$}
    \STATE $\rho^*$ is the solution to $D_2(\rho)=0$
  \ENDIF
\end{algorithmic}
\end{algorithm}

\section{Maximization of Sensor Active Probability in AD-WPT }

In the previous section, we discussed the optimal AD-WPT design in flexible-task WSN scenario where the energy consumption levels or tasks  vary for different SNs.
In some other scenarios, e.g., environmental measurement \cite{environment}  and surveillance monitoring \cite{surveillance} systems, the sensing information from each SN is equally important and mutually irreplaceable.
For example, in a forest fire detection systems \cite{IFsensor}, SNs are randomly deployed in a forest collecting temperature and humidity data independently.
% and reporting the results to the nearest fusion center without much in-network cooperations.
% \cite{forest}
%For example, in a habitat monitoring systems, SNs may work independently in spread areas of a forest collecting ...data for ... \cite{environment}.
In these scenarios, the SNs are assigned with equal sensing tasks with a minimum operational power requirement \cite{minpower}, i.e., an SN is active if the received power is beyond the target energy threshold. To achieve higher sensing diversity, it is important to  allow more SNs operating with sufficient power.
In this section, we analyze the optimal charging radius $\rho^*$ in AD-WPT to maximize the active probability $F_s$ of the SNs.
%In this section, we design the optimal charging radius $\rho^*$ to maximize the percentage of active SNs over all SNs in equal-task WSN, which is equivalent to maximize the active probability $Q_s$ of the typical SN.

% wild animal habitat monitoring
%[18] H. Wang, J. Elson, L. Girod, D. Estrin, K. Yao, ¡° ¡°Target
%classification and localization in habitat monitoring¡±. In Proc. of
%IEEE ICASSP, Hong Kong, 2003.

%military, security and health monitoring systems \cite{IFsensor}, require to work continuously by achieving certain amount of received power in any specific location.
%In this section, we discuss the power outage probability at SN$_0$.
%Given a location realization of the PB network, SN$_0$ suffers power outage if the aggregate received power $P_s$ falls below the target threshold $P_{s}^{th}$. As discussed in Section III-D, the power outage probability $F_s$ is given in \eqref{tau}. In this section, we study the optimal charging radius $\rho^*$ that minimizes the power outage probability.

%\subsection{Optimal Charging Radius}
%As the increase of $\rho$, the power outage probability $\tau_s$ may decrease or increase depending on the values of $\lambda_p$ and $P_p$.
%This is due to the tradeoff between the instantaneous received power from the near PBs $P_{sn}$ and far PBs $P_{fn}$.

As discussed in Section IV,  the decreased number of beams  at the PBs improves the radiated power intensity, which enhances the average received power at SN$_0$ in AD-WPT compared with omnidirectional WPT. However, the decreased number of beams may not enhance the sensor active probability $F_s$ due to the interplay  between the power intensity and power opportunity.
\begin{itemize}
\item \textit{Near PBs}: The near PBs help improve the sensor active probability in AD-WPT compared with that in omnidirectional WPT. Since the near PBs always radiate energy towards SN$_0$ with probability $1$ (see \eqref{probnear}) and antenna  gain greater than $1$ (see \eqref{Gmn}), the received power from the near PBs in AD-WPT is higher than that in omnidirectional WPT. %Moreover, $P_{sn}$ increases as the decreased beamwidth.
    With the decreased number of beams from the near PBs, the power intensity increases, which increases the received power from the near PBs and may improve the sensor active probability in AD-WPT.
\item \textit{Far PBs}:  The far PBs can reduce the sensor active probability in AD-WPT compared with that in omnidirectional WPT. Since the far PBs may not radiate energy towards SN$_0$, the received power from a far PB in AD-WPT is higher than that in omnidirectional WPT or zero if SN$_0$ is inside or outside the beamforming directions of the PB, respectively.
With the decreased number of beams from the far PBs, the power intensity of the far PBs increases (see \eqref{Gm0} and \eqref{Gmn}), but the power opportunity to receive energy from the far PBs at SN$_0$ decreases (see \eqref{probfar1} and \eqref{probfar2}).
%The received power at SN$_0$ from a far PB is zero if SN$_0$ is outside the beamforming directions of the PB.
Since SN$_0$ has a higher chance to fall outside the radiation directions of the far PBs, the received power from the far PBs is more likely to decrease, which may reduce the sensor active probability in AD-WPT.
\end{itemize}
As $\rho$ increases from $0$ to $\infty$, the number of beams of most PBs decreases from $N$ to $1$, and then increases from $1$, $2$, $\cdots$, to $N$. %With the decreased number of beams at the PB, the power intensity of the PBs increases, but SN$_0$ is more likely to fall outside the radiation directions of the far PBs.
With fewer beams, the increased power intensity of the near PBs and decreased power opportunity of the far PBs have positive and negative impacts on the sensor active probability $F_s$, respectively.
%there is a tradeoff between the increased power intensity of the near PBs and the decreased power opportunity of the far PBs.
Whether the near PBs or the far PBs dominate $F_s$ depends on the PB power/density and radius $\rho$. If the PB power/density is low or $\rho$ is large, the near PBs dominate $F_s$ due to the severe power attenuation of the far PBs, and vice versa.

%There is a tradeoff between the power intensity of the near PBs and the power opportunity of the far PBs.
% the number of beams of most PBs decreases from $N$ to $1$. The corresponding power intensity of the PBs increases and the power opportunity of the far PBs decreases. As the further increase of $\rho$, the number of beams of most PBs increases from $1$, $2$, $\cdots$, to $N$. The  corresponding power intensity of the near PBs decreases and the power opportunity of the far PBs increases. There is a tradeoff between the power intensity of the near PBs and the power opportunity of the far PBs.  Whether the near PBs or the far PBs dominate $F_s$ depends on the PB power and density.

\begin{figure}[t!]
    \begin{center}
        \includegraphics[width=0.63\columnwidth]{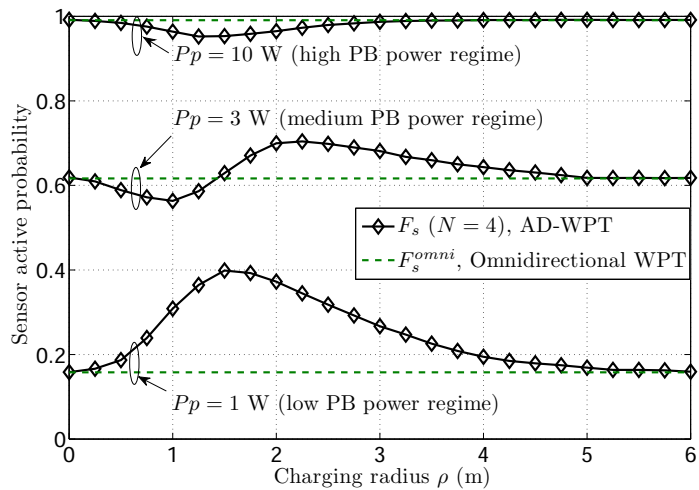}
        \caption{Sensor active probability versus charging radius with various PB power  ($\lambda_p=0.1$, $\lambda_s=0.2$, $P_{s}^{th}=0.1$ mW, $\nu=0.1$ m and $\alpha=3$).}
    \end{center}
\end{figure}

In the following, we analyze the optimal charging radius $\rho^*$ that maximizes the sensor active probability $F_s$ at the typical SN$_0$, i.e.,
\begin{eqnarray}
   \text{P2}:\ &&F_s^*=\max\limits_{0<\rho<{\infty}}{~~~F_s}.
\end{eqnarray}
The simulation results of the sensor active probabilities $F_s$ in AD-WPT  and $F_s^{omni}$ in omnidirectional WPT  are plotted against $\rho$ with various $P_p$ in Fig. 4.
As shown in Corollary 4, the sensor active probability increases with increased PB power $P_p$ or PB density $\lambda_p$.
We discuss $F_s$ by considering different power regimes of the PBs. \\
%In Fig. 4, we show three examples of PB power regimes: low PB power regime (e.g.,  $P_p=1$ W), medium PB power regime (e.g., $P_p=3$ W) and high PB power regime (e.g., $P_p=10$ W).
%For low PB power/density, the near PBs dominate the problem for all $\rho$. For medium PB power/density, the far PBs dominate for small $\rho$ and the near PBs dominate for large $\rho$. For high PB power/density, the far PB dominate the problem for all $\rho$.
%As shown in Fig. 2, the number of beams of most PBs decreases and then increases with the increased $\rho$. \\
\textit{Case 1: Low PB power/density  regime}.
When the PBs have low power $P_p$ and/or density $\lambda_p$, e.g., $P_p=1$ W in Fig. 4, the near PBs dominate $F_s$ and we have $F_s>{F}_s^{omni}$. $F_s$ increases and then decreases with $\rho$ mainly due to the decreased and increased power intensity of the PBs, respectively. There exists an optimal $\rho^*$ (e.g., $\rho=1.5$) that maximizes $F_s$.\\
%In this case, the optimal strategy is to adopt directional WPT with optimal $\rho^*$.
\textit{Case 2: Medium PB power/density  regime}.
When the PBs have medium power $P_p$ and/or density $\lambda_p$, e.g., $P_p=3$ W in Fig. 4, the far PBs dominate $F_s$ for small $\rho$ with ${F}_s\leq{F}_s^{omni}$, and the near PBs dominate $F_s$ for large $\rho$ with ${F}_s>{F}_s^{omni}$, respectively.
There exists an optimal $\rho^*$ (e.g., $\rho=2.25$) that maximizes $F_s$ in the region of ${F}_s>{F}_s^{omni}$.\\
\textit{Case 3: High PB power/density  regime}.
When the PBs have high power $P_p$ and/or density $\lambda_p$, e.g., $P_p=10$ W in Fig. 4, we have  ${F}_s<{F}_s^{omni}$. Due to the high PB power, the sensor active probability ${F}_s^{omni}$ for omnidirectional WPT is high. For AD-WPT, $F_s$ decreases and then increases with $\rho$ mainly due to the decreased and increased power opportunity of the far PBs, respectively.
%For AD-WPT, the fluctuation of the received power from the far PBs reduces the sensor active probability compared with omnidirectional WPT. Since $F_s$ decreases and then increases with $\rho$,
As $\rho\rightarrow{0}$ or $\rho\rightarrow\infty$, we have $F_s\rightarrow{F}_s^{omni}$. The optimal $\rho^*$ is approaching $0$ or $\infty$ as in omnidirectional WPT.

%\textit{Case 5: Infinite Power Supply}.
%When $\lambda_p\rightarrow\infty$ or $P_p\rightarrow\infty$, SN$_0$ is able to harvest enough power from the near PBs and the power outage event is not likely to happen. In this case, we have ${F}_s^{omni}\rightarrow0$ and $\tau_s\rightarrow0$ for all $\rho$.

%From the above discussion, the tradeoff between $P_{s,n}$ and $P_{s,f}$ affect the power reliability. However, it does not influence the average received power. In section IV, it was shown that $E(P_s)$ strictly increases with the decreased beamwidth. This is because that $\mathbb{E}(P_{s,n})$ always increases with the decreased beamwidth and $\mathbb{E}(P_{s,f})$ is equivalent to $\mathbb{E}(P_{f,n}^{omni})$ as discussed in Corollary 2.

%\subsection{Optimal Charging Radius that Minimizes Power Outage Probability}

From the above discussions, we see that the maximized sensor active probability  in AD-WPT with the proper selection of charging radius  $\rho^*$ is larger than or at least equivalent to that in omnidirectional WPT.
%We further derive the optimal charging radius.
As discussed in Section III-D, $F_s$ is not analytically tractable but can be well  approximated using Gamma distribution. In the following, we analyze  the optimal charging radius $\tilde{\rho}^*$ that maximizes the approximation of sensor active probability  $\tilde{F}_s$ in Proposition 4, i.e.,
\begin{eqnarray}
   \text{P3}:\ &&\tilde{F}_s^*=\max\limits_{0<\rho<{\infty}}{~~~\tilde{F}_s}.
\end{eqnarray}
To solve the one-dimensional problem in P3, the optimal radius is one of the stationary points of $\tilde{F}_s$.
Taking the derivative of $\tilde{F}_s$ with respect to $\rho$ yields
%\begin{small}
%\begin{align}
%\frac{\partial\tilde{F}_s}{\partial{\rho}}
%&=-\frac{\left(\frac{P_s^{th}}{\theta}\right)^{k-1}e^{-\frac{P_s^{th}}{\theta}}P_s^{th}\left[\frac{\partial\mathbb{E}(P_s)}{\partial\rho}\mathbb{V}(P_s)+\mathbb{E}(P_s)\frac{\partial\mathbb{V}(P_s)}{\partial\rho}\right]+\left[\frac{\partial\left[\mathbb{E}(P_s)\right]^2}{\partial\rho}\mathbb{V}(P_s)+\left[\mathbb{E}(P_s)\right]^2\frac{\partial\mathbb{V}(P_s)}{\partial\rho}\right]\int_{0}^{\frac{P_s^{th}}{\theta}}t^{k-1}e^{-t}\ln(t)dt}{\left[\mathbb{V}(P_s)\right]^2\Gamma\left(k\right)}\nonumber\\
%&~~~+\frac{\gamma\left(k,\frac{P_s^{th}}{\theta}\right)\left[\frac{\partial\left[\mathbb{E}(P_s)\right]^2}{\partial\rho}\mathbb{V}(P_s)+\left[\mathbb{E}(P_s)\right]^2\frac{\partial\mathbb{V}(P_s)}{\partial\rho}\right]\int_{0}^{\infty}t^{k-1}e^{-t}\ln(t)dt}{\left[\mathbb{V}(P_s)\right]^2\left[\Gamma\left(k\right)\right]^2}\label{Fs},
%\end{align}
%\end{small}
\begin{align}
\frac{\partial\tilde{F}_s}{\partial{\rho}}
&=-\frac{\left(\frac{P_s^{th}}{\theta}\right)^{k-1}e^{-\frac{P_s^{th}}{\theta}}P_s^{th}\left[\frac{\partial\mathbb{E}[P_s]}{\partial\rho}\mathbb{V}[P_s]+\mathbb{E}[P_s]\frac{\partial\mathbb{V}[P_s]}{\partial\rho}\right]}{\left[\mathbb{V}[P_s]\right]^2\Gamma\left(k\right)}\nonumber\\
&~~~-\frac{\left[\frac{\partial\left[\mathbb{E}[P_s]\right]^2}{\partial\rho}\mathbb{V}[P_s]+\left[\mathbb{E}[P_s]\right]^2\frac{\partial\mathbb{V}[P_s]}{\partial\rho}\right]\int_{0}^{\frac{P_s^{th}}{\theta}}t^{k-1}e^{-t}\ln(t)dt}{\left[\mathbb{V}[P_s]\right]^2\Gamma\left(k\right)}\nonumber\\
&~~~+\frac{\gamma\left(k,\frac{P_s^{th}}{\theta}\right)\left[\frac{\partial\left[\mathbb{E}[P_s]\right]^2}{\partial\rho}\mathbb{V}[P_s]+\left[\mathbb{E}[P_s]\right]^2\frac{\partial\mathbb{V}[P_s]}{\partial\rho}\right]\int_{0}^{\infty}t^{k-1}e^{-t}\ln(t)dt}{\left[\mathbb{V}[P_s]\right]^2\left[\Gamma\left(k\right)\right]^2}\label{Fs},
\end{align}
where ${\partial\mathbb{E}[P_s]}/{\partial\rho}$ is given in Section IV and ${\partial\mathbb{V}[P_s]}/{\partial\rho}$ can be obtained via similar approaches.
The number of stationary points of $\tilde{F}_s$ is less than or equal to two since
${\partial\tilde{F}_s}/{\partial{\rho}}=0$ has at most two solutions for $\rho\in(0,\infty)$.
%In Case 1, the optimal radius is the single stationary point that maximizes $\tilde{F}_s$. In Case 2, $\tilde{F}_s$ has two stationary points which minimize and maximize $\tilde{F}_s$, respectively. The optimal radius is the stationary point with larger value of $\rho$. In Case 3,
%$\tilde{F}_s$ has one stationary point that minimizes $\tilde{F}_s$. The optimal radius is approaching $0$ or $\infty$.
%Moreover, we have $\frac{\partial\tilde{F}_s}{\partial{\rho}}|_{\rho=0^+}>0$ for Case 1 and $\frac{\partial\tilde{F}_s}{\partial{\rho}}|_{\rho=0^+}<0$ for Case 2 and 3.
The procedure to obtain the optimal radius for P3 is summarized in Algorithm 2.
\begin{algorithm}[h]
 \caption{Solving the optimal charging radius in P3:}
\begin{algorithmic}[1]
  \STATE Find  $J$ as the number of stationary points of $\tilde{F}_s$  for $\rho\in(0,\infty)$ and calculate $\frac{\partial\tilde{F}_s}{\partial{\rho}}|_{\rho=0^+}$
  \IF{$J=1$ and $\frac{\partial\tilde{F}_s}{\partial{\rho}}|_{\rho=0^+}<0$ (as in Case 1)}
    \STATE $\tilde{\rho}^*$ is the single stationary point of $\tilde{F}_s$
  \ELSIF{$J=2$ and $\frac{\partial\tilde{F}_s}{\partial{\rho}}|_{\rho=0^+}>0$ (as in Case 2)}
    \STATE $\tilde{\rho}^*$ is the stationary point of  $\tilde{F}_s$ with larger value of $\rho$
  \ELSIF{$J=1$ and $\frac{\partial\tilde{F}_s}{\partial{\rho}}|_{\rho=0^+}>0$ (as in Case 3)}
    \STATE $\tilde{\rho}^*$ is  approaching $0$ or $\infty$
  \ENDIF
\end{algorithmic}
\end{algorithm}

%The stationary points of $\tilde{\tau}_s$ are derived by solving $\frac{\partial\tilde{\tau}_s}{\partial{\rho}}=0$.
%Then, by substituting the stationary points back into $\tilde{\tau}_s$, we obtain the optimal approximation of $\tilde{\rho}^*$ that gives the minimum power outage probability.

% mention beamwidth in E(Ps)
% quote V(Ps) for outage
% allocation scheme compare
% cite corollary 4 in fig. 4 Pp tau

\section{Numerical Results}
In this section, we present the simulation results of the maximized average received power for flexible-task WSN and the maximized sensor active probability for equal-task WSN under the proposed AD-WPT scheme, respectively. The performance of omnidirectional WPT scheme is used as a comparison  benchmark. Throughout this section, we set $\sigma=-41.9842$ dB, where the wavelength is $\nu=0.1$ m and reference distance is $d_0=1$ m.

\subsection{Maximized Average Received Power for Flexible-task WSN}

\begin{figure}[t!]
    \begin{center}
        \includegraphics[width=1\columnwidth]{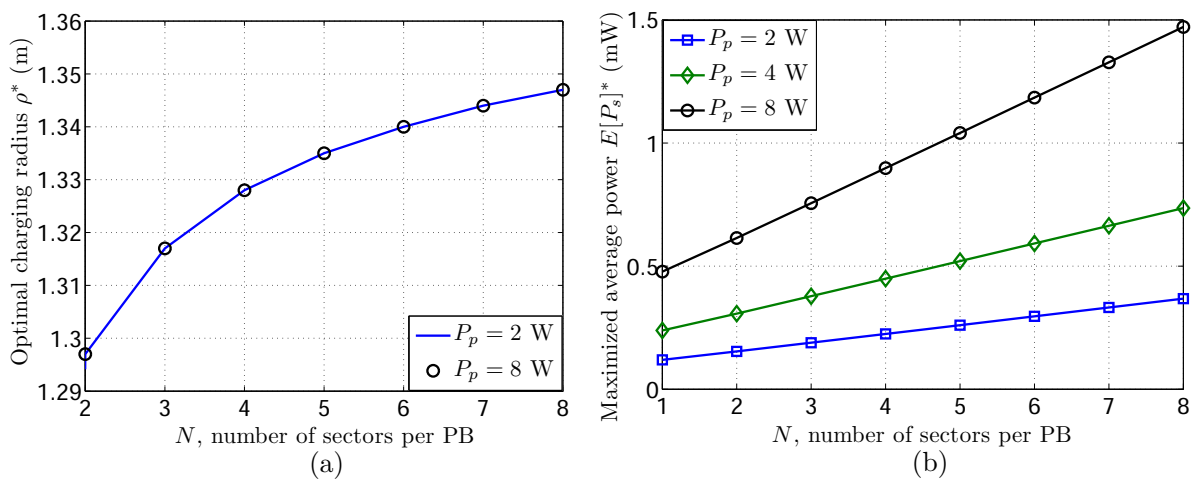}
        \caption{(a) Optimal charging radius $\rho^*$ for average power maximization versus $N$ ($\lambda_p=0.1$, $\lambda_s=0.2$ and $\alpha=3$). (b) Maximized average received power $\mathbb{E}[P_s]^*$ versus $N$ ($\lambda_p=0.1$, $\lambda_s=0.2$ and $\alpha=3$).}
    \end{center}
    \label{rhoN}
\end{figure}

%\begin{figure}[t!]
%    \begin{center}
%        \includegraphics[width=1\columnwidth]{EpsNnew.eps}
%        \caption{Maximized average received power versus $N$ ($\lambda_p=0.1$, $\lambda_s=0.2$ and $\alpha=3$).}
%    \end{center}
%    \label{EpsN}
%\end{figure}

%\begin{figure}[t!]
%    \begin{center}
%        \includegraphics[width=1\columnwidth]{Epslambdasnew.eps}
%        \caption{Maximized average received power versus $\lambda_s$ ($\alpha=3$, $\lambda_p=0.1$ and $N=4$).}
%    \end{center}
%    \label{Epslambdas}
%\end{figure}
Fig. 5 (a) and Fig. 5 (b) show that both the maximized average received power $\mathbb{E}[P_s]^*$ and the corresponding optimal charging radius $\rho^*$ increase with the increased number of PB sectors $N$.
As $N$ increases, the PBs are able to form narrower energy beams with higher power intensity towards the intended SNs. As a result, the coverage of PBs in AD-WPT extends and it is more beneficial to use a larger charging radius $\rho^*$ as shown in Fig. 5 (a) to serve more SNs efficiently.
%The maximized average received power at SN$_0$ also increases with the increased power intensity of the PBs as shown in Fig. 6.
With the decreased beamwidth, the power intensity of the near PBs increases, which thus improves $\mathbb{E}[P_s]^*$ as shown in Fig. 5 (b). This is similar to the effect of decreasing the number of beams of the PBs as discussed in Section IV.

\begin{figure}[t!]
    \begin{center}
        \includegraphics[width=1\columnwidth]{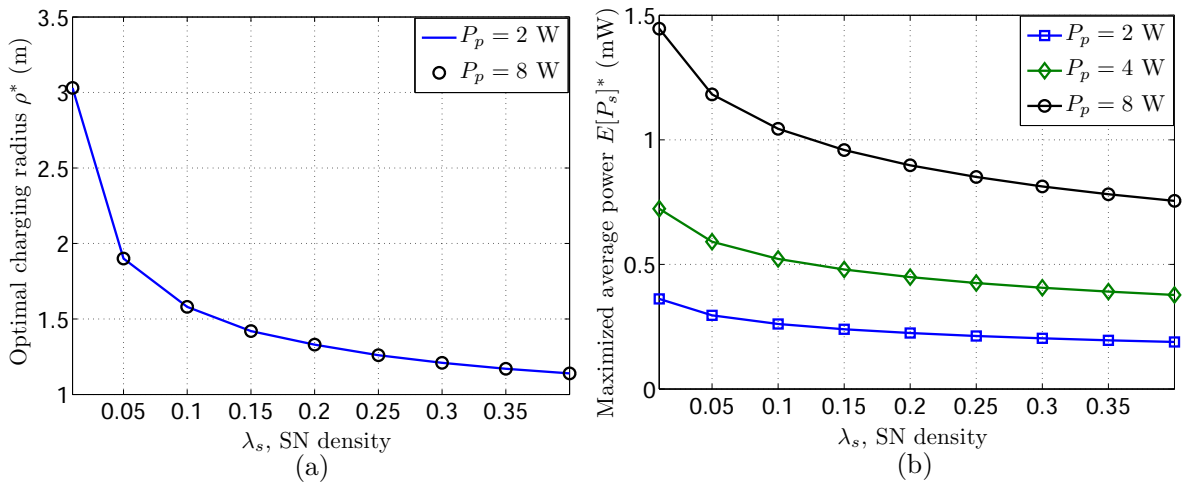}
        \caption{(a) Optimal charging radius $\rho^*$ for average power maximization versus $\lambda_s$ ($\lambda_p=0.1$, $\alpha=3$ and $N=4$). (b) Maximized average received power $\mathbb{E}[P_s]^*$ versus $\lambda_s$ ($\lambda_p=0.1$, $\alpha=3$ and $N=4$).}
    \end{center}
    \label{rholambdas}
\end{figure}

Fig. 6 (a) and Fig. 6 (b) illustrate that the maximized average received power $\mathbb{E}[P_s]^*$ and the corresponding optimal charging radius $\rho^*$  decrease with the increased SN density $\lambda_s $. As $\lambda_s$ increases, more sectors of PBs are activated. The PBs form more energy beams with lower power intensity towards the intended SNs. As a result, the PBs shrink the charging radius in AD-WPT as shown in Fig. 6 (a) to serve fewer SNs efficiently.
%With the decreased optimal radius, the number of near PBs at SN$_0$ decreases.
%These near PBs may still radiate power in large number of beams with low power intensity due to high SN density.
In Fig. 6 (b), $\mathbb{E}[P_s]^*$ decreases with the increased $\lambda_s$ due to the decreased power intensity of the near PBs.

From Fig. 5 (a) to Fig. 6 (b),  we observe that $\mathbb{E}[P_s]^*$ increases linearly with the increased PB power $P_p$, but $\rho^*$ is regardless of $P_p$.
It can also be deduced that increasing $\lambda_p$ has a similar impact on $\mathbb{E}[P_s]^*$ as increasing $P_p$.
Moreover, when $P_p$ is relatively high (e.g., $P_p=8$ W), increasing $N$ or decreasing $\lambda_s$ causes more significant improvement of $\mathbb{E}[P_s]^*$ as shown in Fig. 5 (b) and Fig. 6 (b), respectively.

%Fig. \ref{rhoalpha} and Fig. \ref{Epsalpha} show that $\rho^*$ and $\textsf{E}\left[P_s^{p}(\rho^*)\right]$ decrease as the increase of the path loss exponent $\alpha$. Intuitively, the power transfer efficiency is low in a severe path loss environment.

%\begin{figure}[t!]
%    \begin{center}
%        \includegraphics[width=1\columnwidth]{rhoalpha.eps}
%        \caption{$\rho^*$ vs $\alpha$ ($P_p=1000$ W, $\nu=0.1$ m, $\lambda_p=1$, $\beta=1$, $N=8$). }
%    \label{rhoalpha}
%    \end{center}
%\end{figure}
%\begin{figure}[t!]
%    \begin{center}
%        \includegraphics[width=1\columnwidth]{Epsalpha.eps}
%        \caption{$\textsf{E}\left[P_s^{p}(\rho^*)\right]$ vs $\alpha$ ($P_p=1000$ W, $\nu=0.1$ m, $\lambda_p=1$, $\beta=1$, $N=8$). }
%    \label{Epsalpha}
%    \end{center}
%\end{figure}

%\begin{figure}[t!]
%    \begin{center}
%        \includegraphics[width=1.00\columnwidth]{2000.eps}
        %\caption{$\textsf{E}\left[P_s^{dir}\right]_{max}$ vs $N$ ($P_p=1000$ W, $\nu=0.1$ m, $\alpha=3$, $\lambda_p=1$ and  $\beta=1$).}
%    \end{center}
%    \label{EpsN}
%\end{figure}

%\begin{figure}[t!]
%    \begin{center}
%        \includegraphics[width=1.00\columnwidth]{4000.eps}
        %\caption{$\textsf{E}\left[P_s^{dir}\right]_{max}$ vs $N$ ($P_p=1000$ W, $\nu=0.1$ m, $\alpha=3$, $\lambda_p=1$ and  $\beta=1$).}
%    \end{center}
%    \label{EpsN}
%\end{figure}

\subsection{Maximized Sensor Active Probability  for Equal-task WSN}
%Next, we study the simulation results of the maximized sensor active probability at the optimal charging radius and its relation with the SN density $\lambda_s$ and the number of sectors per PB $N$.

%Fig. 9 shows that the minimized power outage probability decreases with the increased $N$.
In Fig. 7 (a), the maximized sensor active probability $F_s^*$ increases with the increased $N$. As discussed in section V, the improvement of the sensor active probability in AD-WPT compared with omnidirectional WPT is mainly due to the high power intensity of the near PBs.
%As discussed in Fig. 4, the maximized sensor active probability is within the range of $F_s\geq{F}_s^{omni}$ and it is mainly due to the high power intensity of energy beams of the near PBs.
As $N$ increases, the beamwidth of the PBs decreases, which improves the power intensity and thus improves $F_s^*$.
%We notice that the decreasing speed of the minimized sensor inactive probability against $N$ slows down when $P_p$ is large.
%It is because that both $F_s$ and $F_s^{omni}$ are low and the improvement is almost not noticeable.
Fig. 7 (b) shows the  maximized sensor active probability $F_s^*$ decreases with the increased $\lambda_s$.
As $\lambda_s$ increases, the near PBs are likely to form more beams with lower power intensity, which therefore reduces $F_s^*$. Furthermore, we notice that increasing $N$ or decreasing $\lambda_s$ causes more significant improvement of $F_s^*$ for relative small $P_p$ (e.g., $P_p=2$ W). The improvement is less significant for relatively large PB power (e.g., $P_p=8$ W) since the sensor active probability in omnidirectional WPT is already high and may not be much improved by AD-WPT.

\begin{figure}[t!]
    \begin{center}
        \includegraphics[width=1.00\columnwidth]{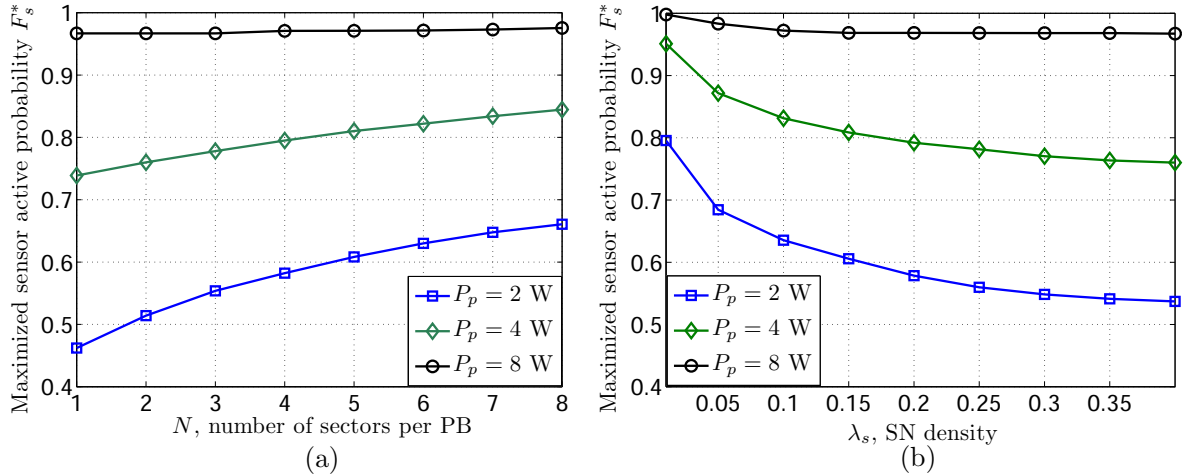}
        \caption{(a) Maximized sensor active probability $F_s^*$ versus $N$  ($\lambda_p=0.1$, $\lambda_s=0.2$, $P_{s}^{th}=0.1$ mW and $\alpha=3$). (b) Maximized sensor active probability $F_s^*$ versus $\lambda_s$ ($\lambda_p=0.1$, $N=4$, $P_{s}^{th}=0.1$ mW and $\alpha=3$).}
    \end{center}
\end{figure}

%\begin{figure}[t!]
%    \begin{center}
%        \includegraphics[width=1.00\columnwidth]{CCDFlambdas.eps}
%        \caption{Maximized sensor active probability versus $\lambda_s$ ($\lambda_p=0.1$, $N=4$, $P_{s}^{th}=0.1$ mW and $\alpha=3$).}
%    \end{center}
%\end{figure}

\subsection{Comparison with Other Power Allocation Schemes}

For the AD-WPT scheme in Section II-A, we adopt uniform power allocation for the PBs, i.e., uniformly allocating the PB power among all active sectors  that have at least one SN.
If the exact number of SNs in each sector is known, the PBs may adopt unequal power allocation schemes which allocate the PB power according to  the number of SNs in each sector.
In this subsection, we mainly discuss two other power allocation schemes: greedy scheme and robust scheme. In greedy scheme, each PB allocates all power to the sector that has the largest number of SNs and no power to all other sectors. It can be easily shown that this scheme provides the maximum sum received power of all SNs in the charging region of a PB. In robust scheme, each PB allocates power proportionally to the number of SNs in each sector.

\begin{figure}[t!]
    \begin{center}
        \includegraphics[width=1.00\columnwidth]{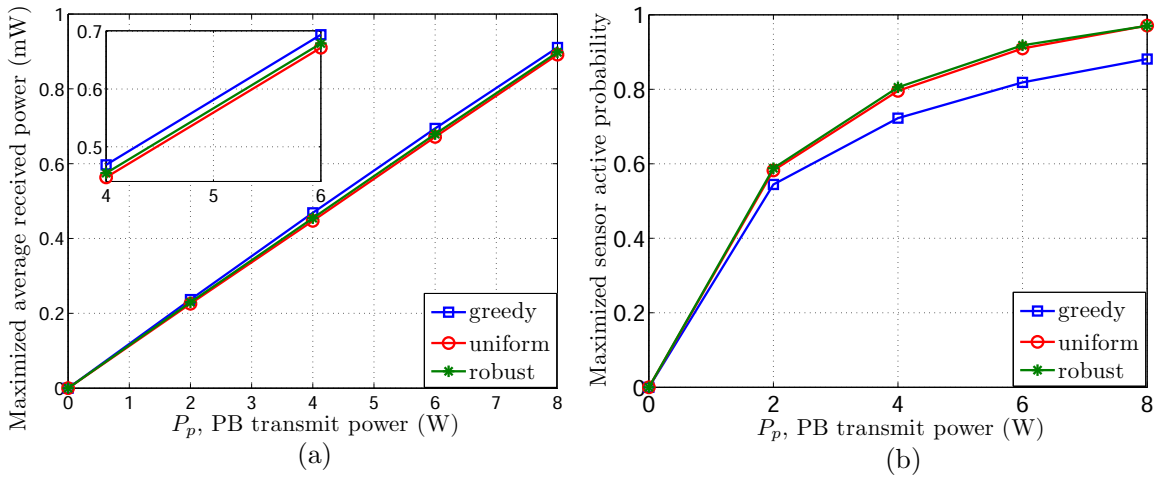}
        \caption{(a) Comparison of the maximized average received power for the three power allocation schemes ($\lambda_p=0.1$, $\lambda_s=0.2$, $N=4$ and $\alpha=3$). (b) Comparison of maximized sensor active probability for the three power allocation schemes ($\lambda_p=0.1$, $\lambda_s=0.2$, $P_s^{th}=0.1$ mW, $N=4$ and $\alpha=3$).}
    \end{center}
\end{figure}

In Fig. 8 (a) and Fig. 8 (b), we compare the maximized average received power and the maximized sensor active probability for the three power allocation schemes, respectively. Fig. 8 (a) shows that the maximized average received power for the three schemes is similar with very minor gaps. Greedy scheme performs the best and robust scheme slightly outperforms uniform scheme in terms of average received power. In Fig. 8 (b), we see that the maximized sensor active probability of robust scheme is the highest, which slightly outperforms that of uniform scheme. The sensor active probability of greedy scheme is the lowest.

To sum up, greedy scheme shows the highest average received power but at the cost of lowest sensor active probability. It is because the single beam strategy in greedy scheme improves the power intensity but also reduces the power opportunity towards SN$_0$. Moreover, robust scheme outperforms uniform scheme in both average power and active probability of the SNs, but the improvement is insignificant. For greedy and robust schemes, the exact number of SNs in each sector is required and the derivation of the distribution of the received power in a heterogeneous network is more complicated than uniform scheme since the gain of the PB becomes a continuous instead of discrete variable. %The derivation of Laplace transform is not a simple decomposition of multiple homogeneous network
%This is because the gain of the PB in adaptive schemes is a continuous value instead of discrete value as in uniform scheme. The decomposition of the heterogeneous network into multiple homogeneous network is no longer applicable.
For uniform scheme, each PB needs only the information of the existence of SNs in each sector. It provides acceptable average power and active probability  with less implementation complexity.

%\begin{figure}[t!]
%    \begin{center}
%        \includegraphics[width=1.00\columnwidth]{CCDFscheme.eps}
%        \caption{Comparison of maximized sensor active probability for the three power allocation schemes ($\lambda_p=0.1$, $\lambda_s=0.2$, $P_s^{th}=0.1$ mW, $N=4$ and $\alpha=3$). }
%    \end{center}
%\end{figure}

\section{Conclusions}
In this paper, we proposed an AD-WPT scheme in a large-scale sensor network, where the PBs charge the SNs by adapting the energy beamforming strategies to the nearby  SN locations.
%The SNs that are inside the power regions of the PBs receive energy from PBs via energy beamforming.
%The SNs that are outisde the power regions of PBs also opportunistically harvest energy from the PBs that radiate power in its direction.
By using stochastic geometry, we derived the closed-form expressions of the distribution metrics, i.e., Laplace transform, mean and variance, of the aggregate received power at a typical SN. The approximation of the CCDF of received power is obtained using Gamma distribution with second-order moment matching. For flexible-task and equal-task WSN, the optimal radii in AD-WPT were designed to maximize the average received power and maximize the sensor active probability, respectively. The results show that the maximized average received power and maximized sensor active probability increase with the increased density and power of the PBs, while they decrease with the increased density of the SNs and energy beamwidth.
Moreover, the optimal AD-WPT is more energy efficient than omnidirectional WPT by achieving equivalent average received power or sensor active probability with less transmit power consumptions.

\section*{Appendix}
\subsection{Proof of Lemma 3}
%In this appendix, we derive the Laplace transform of the distribution of $P_{s,n}^{M}$ in Lemma 3.
Taking the Laplace transform of \eqref{Psnnn}, we have
\begin{align}
\mathcal{L}_{P_{s,n}^{M}}(s)
&=\mathbb{E}\left[\exp\left(-sP_{s,n}^{M}\right)\right]=\mathbb{E}\left[\exp\left[-sP_p\sigma\sum\limits_{X_i\in\Phi_{p,n}^M}{G}_M\left[\max\left(\|X_i\|,1\right)\right]^{-\alpha}\right]\right]\nonumber\\
&=\mathbb{E}\left[\prod\limits_{X_i\in\Phi_{p,n}^M}\exp\left[-sP_p\sigma{G}_M\left[\max\left(\|X_i\|,1\right)\right]^{-\alpha}\right]\right]\nonumber\\
&=\exp\left[-\lambda_p\eta_n^M\int_{0}^{2\pi}\int_{0}^{\rho}\left[1-\exp\left[-sP_p\sigma{G}_M\left[\max\left(r,1\right)\right]^{-\alpha}\right]\right]rd\psi{d}r\right].\label{LPsnM}
\end{align}
The last step is obtained by applying probability generating functional (PGFL)~\cite[Proposition 2.12]{weber}, where $r$ and $\psi$
denote the radial coordinate and angular coordinate in polar coordinate system.
%where $\psi$ and $r$ are the
For $0<\rho\leq1$, \eqref{LPsnM} is further derived as
\begin{align}
\mathcal{L}_{P_{s,n}^{M}}(s)&=\exp\left[-2\pi\lambda_p\eta_n^M\int_{0}^{\rho}\left[1-\exp\left(-sP_p\sigma{G}_M\right)\right]rdr\right].\label{LP1}
%&=\exp\Bigg\{-2\pi\lambda_p\eta_n^M\left[\frac{1}{2}\rho^2-\frac{1}{2}\rho^2\exp\left[-sP_p\left(\frac{\nu}{4\pi}\right)^2{G}_M\right]\right]\Bigg\}\nonumber\\
%&=\exp\left[-\lambda_p\pi{\eta}_{n}^{M}\left[\rho^2-\rho^2\exp\left({-s{P}_p\sigma{G}_M}\right)\right]\right]\nonumber.
\end{align}
For $1<\rho<\infty$, \eqref{LPsnM} is further derived as
\begin{align}
\mathcal{L}_{P_{s,n}^{M}}(s)
&=\exp\left[-2\pi\lambda_p\eta_n^M\left[\int_{0}^{1}\left[1-\exp\left(-sP_p\sigma{G}_M\right)\right]rdr\right.\right]\nonumber\\
&~~~~~~~~~~~~~~~~~~~~~~~~~\left.\left.-\int_{1}^{\rho}\left[1-\exp\left(-sP_p\sigma{G}_Mr^{-\alpha}\right)\right]rdr\right]\right].\label{LP2}
%&=\exp\left[-\lambda_p\pi{\eta}_{n}^{M}\left[\rho^2-\rho^2\exp\left({-s{P}_p\sigma{G}_M{\rho}^{-\alpha}}\right)\right.\right.\nonumber\\
%&~~~~~~~~~~~+\left.\left.\left(s{P}_p\sigma{G}_M\right)^{\frac{2}{\alpha}}\left[\gamma\left(1-\frac{2}{\alpha},s{P}_p\sigma{G}_M\right)-\gamma\left(1-\frac{2}{\alpha},s{P}_p\sigma{G}_M\rho^{-\alpha}\right)\right]\right]\right].\nonumber
\end{align}
From \eqref{LP1} and \eqref{LP2}, we can easily obtain \eqref{LL1} and \eqref{LL2} in Lemma 3, respectively.

%\begin{align}
%\int{x}^me^{-\beta{x}^n}dx=-\frac{\Gamma\left[\gamma,\beta{x}^n\right]}{n\beta^{\gamma}}=-\frac{\int_{\beta{x}^n}^{\infty}t^{\gamma-1}e^{-t}dx}{n\beta^{\gamma}}
%\end{align}
%where $\gamma=\frac{m+1}{n}$

\subsection{Proof of Lemma 4}
%In this appendix, we derive the Laplace transform of the distribution of $P_{s,f}^{M}$ in Lemma 4. Similar to \eqref{LPsnM}, we have
Taking the Laplace transform of \eqref{Psfff}, we have
\begin{align}
\mathcal{L}_{P_{s,f}^{M}}(s)
&=\mathbb{E}\left[\exp\left(-sP_{s,f}^{M}\right)\right]\nonumber\\
&=\exp\Bigg\{-\lambda_p\eta_f^M\int_{0}^{2\pi}\int_{\rho}^{\infty}\left[1-\exp\left[-sP_p\sigma{G}_M\left[\max\left(r,1\right)\right]^{-\alpha}\right]\right]rd\psi{d}r\Bigg\}\label{LPsfM}.
\end{align}
For $0<\rho\leq1$, \eqref{LPsfM} is derived as
\begin{align}
\mathcal{L}_{P_{s,f}^{M}}(s)
&=\exp\Bigg\{-2\pi\lambda_p\eta_f^M\int_{\rho}^{1}\left[1-\exp\left(-sP_p\sigma{G}_M\right)\right]rdr\nonumber\\
&~~~~~~~~~~-2\pi\lambda_p\eta_f^M\int_{1}^{\infty}\left[1-\exp\left(-sP_p\sigma{G}_Mr^{-\alpha}\right)\right]rdr\Bigg\}\label{LP3}.
\end{align}
For $1<\rho<\infty$, \eqref{LPsfM} is  derived as
\begin{align}
\mathcal{L}_{P_{s,f}^{M}}(s)
=\exp\Bigg\{-2\pi\lambda_p\eta_f^M\int_{\rho}^{\infty}\left[1-\exp\left(-sP_p\sigma{G}_Mr^{-\alpha}\right)\right]rdr\Bigg\}\label{LP4}.
\end{align}
From \eqref{LP3} and \eqref{LP4}, we can easily obtain \eqref{LL3} and \eqref{LL4} in Lemma 4, respectively.

\subsection{Proof of Proposition 2}
In this appendix, we derive $\mathbb{E}(P_s)$ in Proposition 2.
\subsubsection{$0<\rho\leq1$}
%From Proposition 1, the Laplace transform of $P_s$ for $0<\rho\leq1$ is given by
%\begin{align}
%&\mathcal{L}_{P_{s}}(s)|_{0<\rho\leq1}
%=\exp\Bigg\{{\lambda_p\pi}\rho^2\left(\sum\limits_{M=0}^N\eta_f^M-\sum\limits_{M=1}^N\eta_n^M\right)+\lambda_p\pi\rho^2\sum\limits_{M=1}^N\eta_n^M\exp\left({-s{P}_p\sigma{G}_M}\right)\nonumber\\
%&-{\lambda_p\pi}\rho^2\sum\limits_{M=0}^N\eta_f^M\exp\left({-s{P}_p\sigma{G}_M}\right)-{\lambda_p\pi}\sum\limits_{M=0}^N\eta_f^M\left({s{P}_p\sigma{G}_M}\right)^{\frac{2}{\alpha}}\gamma\left(1-\frac{2}{\alpha},s{P}_p\sigma{G}_M\right)\Bigg\}\label{LLL}
%\end{align}
Taking the first derivative of the Laplace transform in Proposition 1 for $0<\rho\leq1$, we have
\begin{align}
&\mathbb{E}[P_s]|_{0<\rho\leq1}
=-\frac{d}{ds}\left[\log\left(\mathcal{L}_{P_{s}}(s)|_{0<\rho\leq1}\right)\right]|_{s=0}\nonumber\\
&=-\lambda_p\pi\rho^2\sum\limits_{M=1}^N\eta_n^M\frac{d}{ds}\left[\exp\left({-s{P}_p\sigma{G}_M}\right)\right]|_{s=0}
+{\lambda_p\pi}\rho^2\sum\limits_{M=0}^N\eta_f^M\frac{d}{ds}\left[\exp\left({-s{P}_p\sigma{G}_M}\right)\right]|_{s=0}\nonumber\\
&~~~+{\lambda_p\pi}\sum\limits_{M=0}^N\eta_f^M\frac{d}{ds}\left[\left({s{P}_p\sigma{G}_M}\right)^{\frac{2}{\alpha}}\gamma\left(1-\frac{2}{\alpha},s{P}_p\sigma{G}_M\right)\right]|_{s=0}.\label{EE1}
\end{align}
By further derivation, we have
\begin{align}
\frac{d}{ds}\left[\exp\left(-{{P}_p\sigma{G}_Ms}\right)\right]|_{s=0}
%&=-{P}_p\left(\frac{\nu}{4\pi}\right)^2\frac{N}{M}{\beta}^{-\alpha}\exp\left(-{{P}_p\left(\frac{\nu}{4\pi}\right)^2\frac{N}{M}s{\beta}^{-\alpha}}\right)\nonumber\\
=-{P}_p\sigma{G}_M\label{G11}
\end{align}
and
%\begin{align}
%\frac{\partial{\gamma\left(a,b\right)}}{\partial{b}}=b^{a-1}e^{-b}
%\end{align}
%Then, we have
\begin{align}
&\lim\limits_{s\rightarrow{0}}\frac{d}{ds}\left[\left({P}_p\sigma{G}_Ms\right)^{\frac{2}{\alpha}}\gamma\left(1-\frac{2}{\alpha},{P}_p\sigma{G}_Ms\right)\right]
=\frac{{P}_p\sigma{G}_M\alpha}{\alpha-2}.\label{G21}
\end{align}
Substituting \eqref{G11} and \eqref{G21} into \eqref{EE1} yields
\begin{align}
&\mathbb{E}[P_s]|_{0<\rho\leq1}
={\lambda_p{P}_p}{\pi}\sigma\left[\rho^2\left(\sum\limits_{M=1}^N\eta_n^MG_M-\sum\limits_{M=0}^N\eta_f^MG_M\right)+\frac{\alpha}{\alpha-2}\sum\limits_{M=0}^N\eta_f^M{G}_M\right]\label{Ep1}.
\end{align}
We further obtain
\begin{align}
\sum\limits_{M=1}^N\eta_n^MG_M=\frac{1-p^N}{1-p}\label{pn1}
\end{align}
and
\begin{align}
\sum\limits_{M=0}^N\eta_f^MG_M=(p+q)^M=1.\label{pn2}
\end{align}
%$\sum\limits_{M=1}^N\eta_n^MG_M=\frac{1-p^N}{1-p}$
%and
%$\sum\limits_{M=0}^N\eta_f^MG_M=(p+q)^M=1$.
Substituting \eqref{pn1} and \eqref{pn2} back into \eqref{Ep1}, we obtain \eqref{E1} in Proposition 2.
%By substituting \eqref{pn1} and \eqref{pn2} into \eqref{Ep1}, we have
%\begin{align}
%\mathbb{E}(P_s)|_{0<\rho\leq1}={P_p\lambda_p}{\pi}\sigma\left[\rho^2\frac{p-p^N}{1-p}+\frac{\alpha}{\alpha-2}\right].
%\end{align}
%Similarly, we can obtain $\mathbb{E}(P_s)|_{1<\rho<\infty}$ as in \eqref{E2}.

\subsubsection{$1<\rho<\infty$}

Taking the first derivative of the Laplace transform in Proposition 1 for $1<\rho<\infty$, we have
\begin{align}
\mathbb{E}[P_s]|_{1<\rho<\infty}
&=-\frac{d}{ds}\left[\log\left(\mathcal{L}_{P_{s}}(s)|_{1<\rho<\infty}\right)\right]|_{s=0}\nonumber\\
&={P}_p{\lambda_p\pi}\sigma\left[\frac{\alpha}{\alpha-2}\sum\limits_{M=1}^N\eta_n^M{G}_M+\frac{2\rho^{2-\alpha}}{\alpha-2}\left(\sum\limits_{M=0}^N\eta_f^M{G}_M-\sum\limits_{M=1}^N\eta_n^M{G}_M\right)\right].\label{Ep2}
\end{align}
Substituting \eqref{pn1} and \eqref{pn2} into \eqref{Ep2}, we obtain  \eqref{E2} in Proposition 2.

\subsection{Proof of Corollary 1}
We compare \eqref{Eiso} with \eqref{E1} and \eqref{E2}, respectively.
Firstly, \eqref{E1} is compared with \eqref{Eiso}.
\begin{align}
\mathbb{E}\left[P_s\right]|_{0<\rho\leq1}-{\mathbb{E}\left[P_s^{omni}\right]}=P_p\lambda_p\pi\sigma{\rho^2}\frac{p-p^N}{1-p},
\end{align}
where $p$ was given in \eqref{p}.
For $\rho\rightarrow{0}$, it has $p\rightarrow1$ and $\mathbb{E}\left[P_s\right]|_{0<\rho\leq1}\rightarrow{\mathbb{E}\left[P_s^{omni}\right]}$.
For $0<\rho\leq1$, it has $p^N<p<1$ which leads to $\mathbb{E}\left[P_s\right]|_{0<\rho\leq1}>{\mathbb{E}\left[P_s^{omni}\right]}$.

Secondly, \eqref{E2} is compared with \eqref{Eiso}.
\begin{align}
\mathbb{E}\left[P_s\right]|_{1<\rho<\infty}-{\mathbb{E}\left(P_s^{omni}\right)}
%&=P_p\lambda_p\pi\sigma\left[\frac{\alpha\beta^{2-\alpha}-2{\rho}^{2-\alpha}}{\pi(\alpha-2)}\frac{1-p^N}{1-p}+\frac{2{\rho}^{2-\alpha}}{\pi(\alpha-2)}
%-{\frac{\alpha{\beta}^{2-\alpha}}{\pi(\alpha-2)}}\right]\nonumber\\
%&=P_p\lambda_p\pi\sigma\frac{\alpha\beta^{2-\alpha}-2{\rho}^{2-\alpha}}{\pi(\alpha-2)}\left(\frac{1-p^N}{1-p}-1\right)\nonumber\\
&=P_p\lambda_p\pi\sigma\frac{\alpha-2{\rho}^{2-\alpha}}{\alpha-2}\frac{p-p^N}{1-p}.
\end{align}
For $\rho\rightarrow\infty$, it has $p\rightarrow0$ and $\mathbb{E}\left[P_s\right]|_{1<\rho<\infty}\rightarrow{\mathbb{E}\left[P_s^{omni}\right]}$.
For $1<\rho<\infty$, it has $p^N<p<1$. Since $\alpha>2$, we further have ${\rho}^{2-\alpha}<1$ and $\alpha>2{\rho}^{2-\alpha}$. Then, it is proved that $\mathbb{E}\left[P_s\right]|_{1<\rho<\infty}>{\mathbb{E}\left[P_s^{omni}\right]}$.

\subsection{Proof of Proposition 3}
%In this appendix, we derive $\mathbb{V}(P_s)$ in Proposition 3.

%\subsection{$0<\rho\leq1$}
Taking the second derivative of the Laplace transform in Proposition 1, we have
\begin{align}
&\mathbb{V}[P_s]|_{0<\rho\leq1}
=\frac{d^2}{d{s}}\left[\log\left(\mathcal{L}_{P_{s}}(s)|_{0<\rho\leq1}\right)\right]|_{s=0}\nonumber\\
&=\lambda_p\pi\rho^2\sum\limits_{M=1}^N\eta_n^M\frac{d^2}{ds^2}\left[\exp\left({-s{P}_p\sigma{G}_M}\right)\right]|_{s=0}
-{\lambda_p\pi}\rho^2\sum\limits_{M=0}^N\eta_f^M\frac{d^2}{ds^2}\left[\exp\left({-s{P}_p\sigma{G}_M}\right)\right]|_{s=0}\nonumber\\
&~~~-{\lambda_p\pi}\sum\limits_{M=0}^N\eta_f^M\frac{d^2}{ds^2}\left[\left({s{P}_p\sigma{G}_M}\right)^{\frac{2}{\alpha}}\gamma\left(1-\frac{2}{\alpha},s{P}_p\sigma{G}_M\right)\right]|_{s=0}.\label{VV1}
\end{align}
By further derivation, we have
\begin{align}
\frac{d^2}{ds^2}\left[\exp\left(-{{P}_p\sigma{G}_Ms}\right)\right]|_{s=0}=\left({P}_p\sigma{G}_M\right)^2\label{VG1}
\end{align}
and
%\begin{align}
%\frac{\partial{\gamma\left(a,b\right)}}{\partial{b}}=b^{a-1}e^{-b}
%\end{align}
%Then, we have
\begin{align}
&\lim\limits_{s\rightarrow{0}}\frac{d^2}{ds^2}\left[\left({P}_p\sigma{G}_Ms\right)^{\frac{2}{\alpha}}\gamma\left(1-\frac{2}{\alpha},{P}_p\sigma{G}_Ms\right)\right]=-\frac{\alpha}{\alpha-1}\left({P}_p\sigma{G}_M\right)^2.\label{VG2}
\end{align}
Substituting \eqref{VG1} and \eqref{VG2} into \eqref{VV1} yields \eqref{V1}. Similarly, we can obtain $\mathbb{V}[P_s]|_{1<\rho<\infty}$ in \eqref{V2}.

\end{document}